%% file: neurips_2025.tex
\documentclass{article}

\usepackage[final]{neurips_2025}

\usepackage[utf8]{inputenc} 
\usepackage[T1]{fontenc}    
\usepackage{hyperref}       
\usepackage{url}            
\usepackage{booktabs}       
\usepackage{amsfonts}       
\usepackage{nicefrac}       
\usepackage{microtype}      
\usepackage{xcolor}         

\input{preamble}

\title{Efficient Training-Free Online Routing for High-Volume Multi-LLM Serving}

\author{%
  Fangzhou Wu \\
    University of Wisconsin--Madison \\
    \texttt{fwu89@wisc.edu}
    \And
    Sandeep Silwal \\
    University of Wisconsin--Madison \\
    \texttt{silwal@cs.wisc.edu} 
}

\begin{document}

\maketitle

\input{abs}

\input{introduction}

\input{related}

\input{problem}

\input{algo}

\input{proof}

\input{experiments}

\input{discussion}

\input{conclusion}

\bibliographystyle{plain}
\bibliography{references}




\newpage
\section*{NeurIPS Paper Checklist}

\begin{enumerate}

\item {\bf Claims}
    \item[] Question: Do the main claims made in the abstract and introduction accurately reflect the paper's contributions and scope?
    \item[] Answer: \answerYes{} 
    \item[] Justification: Yes, the method, theoretical analysis, and experiment sections justify all the claims made in the abstract and introduction. Additional clarifications are also provided in the Appendix.

\item {\bf Limitations}
    \item[] Question: Does the paper discuss the limitations of the work performed by the authors?
    \item[] Answer: \answerYes{} 
    \item[] Justification: Yes, we outline the limitation of our method in~\Cref{sec:conclusion}.

\item {\bf Theory assumptions and proofs}
    \item[] Question: For each theoretical result, does the paper provide the full set of assumptions and a complete (and correct) proof?
    \item[] Answer: \answerYes{} 
    \item[] Justification: Yes, we provide the full set of assumptions, lemmas, and the main theorem in~\Cref{sec:thm}, along with detailed discussions and complete proofs in~\Cref{app:proof}.

    \item {\bf Experimental result reproducibility}
    \item[] Question: Does the paper fully disclose all the information needed to reproduce the main experimental results of the paper to the extent that it affects the main claims and/or conclusions of the paper (regardless of whether the code and data are provided or not)?
    \item[] Answer: \answerYes{} 
    \item[] Justification: Yes, we disclose all information needed to reproduce the main experimental results
of the paper in~\Cref{sec:eval} and~\Cref{app:exp}.

\item {\bf Open access to data and code}
    \item[] Question: Does the paper provide open access to the data and code, with sufficient instructions to faithfully reproduce the main experimental results, as described in supplemental material?
    \item[] Answer: \answerYes{} 
    \item[] Justification: Yes, all the models and datasets we work with in the paper are open-source, and we have submitted our code in the Supplementary Material.

\item {\bf Experimental setting/details}
    \item[] Question: Does the paper specify all the training and test details (e.g., data splits, hyperparameters, how they were chosen, type of optimizer, etc.) necessary to understand the results?
    \item[] Answer: \answerYes{} 
    \item[] Justification: Yes, all experimental details are clarified in~\Cref{sec:eval} and~\Cref{app:exp}.

\item {\bf Experiment statistical significance}
    \item[] Question: Does the paper report error bars suitably and correctly defined or other appropriate information about the statistical significance of the experiments?
    \item[] Answer: \answerYes{} 
    \item[] Justification: Yes, we report error bars in all experiments involving randomness, such as those with randomized query arrival orders, to capture the variability across runs and evaluate the statistical robustness of the results.

\item {\bf Experiments compute resources}
    \item[] Question: For each experiment, does the paper provide sufficient information on the computer resources (type of compute workers, memory, time of execution) needed to reproduce the experiments?
    \item[] Answer: \answerYes{} 
    \item[] Justification: We provide the computer resources needed for all experiments in~\Cref{app:exp}.
    
\item {\bf Code of ethics}
    \item[] Question: Does the research conducted in the paper conform, in every respect, with the NeurIPS Code of Ethics \url{https://neurips.cc/public/EthicsGuidelines}?
    \item[] Answer: \answerYes{} 
    \item[] Justification: The research conforms with the NeurIPS Code of Ethics.

\item {\bf Broader impacts}
    \item[] Question: Does the paper discuss both potential positive societal impacts and negative societal impacts of the work performed?
    \item[] Answer: \answerYes{} 
    \item[] Justification: Yes, the broader impacts of the work are discussed in~\Cref{app:imp}.
    
\item {\bf Safeguards}
    \item[] Question: Does the paper describe safeguards that have been put in place for responsible release of data or models that have a high risk for misuse (e.g., pretrained language models, image generators, or scraped datasets)?
    \item[] Answer: \answerNA{} 
    \item[] Justification: Our paper does not pose any such risks.

\item {\bf Licenses for existing assets}
    \item[] Question: Are the creators or original owners of assets (e.g., code, data, models), used in the paper, properly credited and are the license and terms of use explicitly mentioned and properly respected?
    \item[] Answer: \answerYes{} 
    \item[] Justification: All the existing models and datasets are appropriately cited.

\item {\bf New assets}
    \item[] Question: Are new assets introduced in the paper well documented and is the documentation provided alongside the assets?
    \item[] Answer: \answerYes{} 
    \item[] Justification: We introduce a new algorithm in the paper and provide its implementation along with detailed documentation to support reproducibility.

\item {\bf Crowdsourcing and research with human subjects}
    \item[] Question: For crowdsourcing experiments and research with human subjects, does the paper include the full text of instructions given to participants and screenshots, if applicable, as well as details about compensation (if any)? 
    \item[] Answer: \answerNA{} 
    \item[] Justification: The paper does not involve crowdsourcing nor research with human subjects.

\item {\bf Institutional review board (IRB) approvals or equivalent for research with human subjects}
    \item[] Question: Does the paper describe potential risks incurred by study participants, whether such risks were disclosed to the subjects, and whether Institutional Review Board (IRB) approvals (or an equivalent approval/review based on the requirements of your country or institution) were obtained?
    \item[] Answer:\answerNA{} 
    \item[] Justification: The paper does not involve crowdsourcing nor research with human subjects.

\item {\bf Declaration of LLM usage}
    \item[] Question: Does the paper describe the usage of LLMs if it is an important, original, or non-standard component of the core methods in this research? Note that if the LLM is used only for writing, editing, or formatting purposes and does not impact the core methodology, scientific rigorousness, or originality of the research, declaration is not required.
    \item[] Answer: \answerNA{} 
    \item[] Justification: This research does not involve LLMs as any important, original, or non-standard components.
\end{enumerate}

\newpage
\appendix
\input{appendix}

\end{document}

%% file: preamble.tex
\usepackage[utf8]{inputenc} 
\usepackage[T1]{fontenc}    
\usepackage{hyperref}       
\usepackage{url}            
\usepackage{booktabs}       
\usepackage{amsfonts}       
\usepackage{nicefrac}       
\usepackage{microtype}      
\usepackage{lipsum}
\usepackage{fancyhdr}       
\usepackage{graphicx}       

\usepackage{xcolor}
\usepackage{tcolorbox}
\tcbuselibrary{breakable}
\usepackage{mathtools}
\usepackage{hyperref} 
\usepackage{cleveref}
\usepackage{algorithm}
\usepackage[noend]{algorithmic}
\usepackage{graphicx}
\usepackage{textcomp}
\usepackage{multirow}
\usepackage{xspace}
\usepackage{listings}
\usepackage{caption}
\usepackage{amsthm}
\usepackage{float}
\usepackage{soul}
\usepackage{upquote}
\usepackage{listings}
\usepackage{enumitem}
\usepackage{extarrows}
\usepackage{stmaryrd}
\usepackage{bbm}
\usepackage{subcaption}
\usepackage{makecell}
\usepackage{array}
\usepackage{dsfont}
\usepackage{colortbl}
\usepackage{tikz}
\usepackage{wrapfig}

\DeclareMathOperator*{\argmax}{argmax}

\theoremstyle{plain}
\newtheorem{theorem}{Theorem}
\newtheorem{lemma}[theorem]{Lemma}
\newtheorem{assumption}{Assumption}

\theoremstyle{definition}
\newtheorem{definition}{Definition}

\crefname{assumption}{assumption}{assumptions}
\Crefname{assumption}{Assumption}{Assumptions}

\newcommand{\est}{C_{est}}
\newcommand{\alg}{C_{alg}}
\newcommand{\eest}{{\hat{C}}}
\newcommand{\opt}{C_{opt}}
\newcommand{\eopt}{\hat{C}_{opt}}
\newcommand{\eps}{\hat{d}_{ij}}
\newcommand{\ops}{{d}_{ij}}

%% file: abs.tex
\begin{abstract}
Increasing demand for Large Language Models (LLMs) services imposes substantial deployment and computation costs on providers. 
LLM routing offers a cost-efficient solution by directing queries to the optimal LLM based on model and query features.
However, existing works primarily focus on offline scenarios and struggle to adapt to online settings with high query volume and constrained token budgets.
In this work, we introduce the first training-free algorithm for online routing scenarios.
Our algorithm leverages approximate nearest neighbor search to efficiently estimate query features and performs a one-time optimization over a small set of initial queries to learn a routing strategy that guides future routing.
We provide theoretical guarantees demonstrating that our algorithm achieves a competitive ratio of $1 - o(1)$ under natural assumptions, which is further validated by extensive experiments across 3 benchmark datasets and 8 baselines, showing an average improvement of \textbf{3.55$\times$} in overall performance, \textbf{1.85$\times$} in cost efficiency, and nearly \textbf{4.25$\times$} in throughput.
Our code is available at \url{https://github.com/fzwark/PORT}.
\end{abstract}

%% file: introduction.tex
\section{Introduction}
The ability of Large Language Models (LLMs) to effectively interpret diverse domain knowledge has rapidly transformed the landscape of automated information processing~\cite{frieder2023mathematical, lehnert2023ai, zhang2025scientific}. 
However, the growing volume of user queries imposes substantial deployment costs on LLM-serving providers~\cite{ding2024hybrid}.
For instance, OpenAI reportedly handles up to 12k user queries per second at peak load~\cite{singh2025_chatgptstats}, while its first Azure supercomputer hosted only around 10k GPUs~\cite{microsoft2020_10kgpu}, pushing it to rush in more resources to prevent cost spikes~\cite{wiggers2025_outofgpu}. 
As a result, improving the overall quality of service, particularly under limited token budget constraints, has become a critical priority for LLM-serving systems.

One straightforward approach is to route queries with different features to different LLMs while balancing cost and performance~\cite{lu2023routing, hu2024routerbench, ong2025routellmlearningroutellms}. 
This idea is based on the observation that different LLMs excel in different domains and have varying cost~\cite{ong2025routellmlearningroutellms}, perfectly aligning with the goal of cost-efficient serving.
Existing works typically rely on the model-based predictors~\cite{ong2025routellmlearningroutellms, somerstep2025carrot} or computationally complex calculations (e.g., KNN, similarity-weighted ranking)~\cite{ong2025routellmlearningroutellms, hu2024routerbench} to predict the optimal LLM to route queries to.
While effective in offline scenarios, these methods face significant limitations in practical online routing.
They are computationally demanding and introduce additional latency, making them impractical for high-volume, low-latency online environments~\cite{ong2025routellmlearningroutellms, hu2024routerbench}. 
They do not easily generalize to dynamic LLM deployments, where any configuration change incurs costly retraining overhead~\cite{somerstep2025carrot, feng2024graphrouter,wang2025mixllm}. 
They are also fundamentally limited in achieving effective online routing under constrained token budgets, as they operate under offline assumptions without accounting for the sequential and uncertain nature of real-world query arrivals~\cite{rama2025cerebrum, vsakota2024fly}.

Our work instead proposes the first theoretically grounded online algorithm tailored for high-volume query routing under limited token budget constraints.
For each query, we employ Approximate Nearest Neighbor Search (ANNS) ~\cite{jayaram2019diskann, malkov2018efficient} to efficiently estimate its features (performance and cost) for each deployed LLM using a historical dataset. 
This dataset can be easily collected and maintained by LLM-serving providers from previously served user queries with diverse query types~\cite{NEURIPS2023_91f18a12}, introduces negligible additional deployment overhead, and presents strong adaptivity to different LLM deployments. 
The routing problem (see \Cref{sec:background} for the formal setup) can be naturally formalized as Mixed-Integer Linear Programming (MILP) that aims to maximize overall performance under token budget constraints.
However, solving the global MILP is infeasible in online settings where queries do not arrive simultaneously.

To achieve efficient and near-optimal routing in the realistic sequential setting, we leverage a key observation by looking at the dual problem: At optimality, the dual objective can be fully parameterized by a single dual variable, which directly yields the optimal routing rule.
This further motivates us to treat this dual variable as a set of \emph{learnable weights} over LLMs, with the parameterized dual objective serving as the optimization target.
Instead of solving the infeasible global dual problem, we adapt it to a partial optimization over a small set of initial observed queries to estimate these weights, which are then applied to guide the routing decisions for subsequent queries.
To improve the generalizability of learned weights to future queries, we adopt a random routing strategy for the observed queries, inspired by ideas from PAC learning~\cite{mehta2007adwords, devanur2009adwords, denis1998pac}.
In addition, we introduce a control parameter into the MILP objective that does not affect the optimal solution structure but helps to limit performance deviation on future queries.

Regarding efficiency, our algorithm only performs a one-time optimization over a small sample set (i.e., observed queries, typically $\approx$ 250) and executes practical ANNS per query, making it significantly more computationally efficient than existing methods and well-suited for online routing scenarios.
In terms of effectiveness, our theoretical analysis guarantees that the proposed algorithm achieves a competitive ratio of $1 - o(1)$ under mild assumptions. 
This is further supported by extensive experiments, where our method outperforms all 8 baseline methods across diverse routing settings and three benchmark datasets, achieving an average improvement of \textbf{3.55$\times$} in performance, \textbf{1.85$\times$} in cost efficiency, and nearly \textbf{4.25$\times$} in throughput. 

To summarize, our main contributions are:  

\begin{enumerate}[leftmargin=*, noitemsep]
    \item We propose the first \textbf{training-free online routing algorithm} tailored for high-volume multi-LLM serving under constrained token budgets.
    \item  Our algorithm estimates query features via efficient ANNS methods, ensuring \textbf{computational scalability}.
    \item  Unlike prior work, our method operates directly on the auxiliary dataset without any model training, introducing {negligible deployment overhead} and enabling \textbf{deployment scalability} across dynamic LLM deployment configurations.
    \item  Our algorithm performs a one-time optimization over a small sample set without requiring intensive computational resources, demonstrating \textbf{high efficiency}. 
    \item  We provide formal theoretical guarantees showing that our algorithm achieves a {competitive ratio of $1 - o(1)$} relative to the offline optimum (which is unknown) under mild assumptions. 
    \item  We conduct extensive experiments on 3 benchmarks against 8 baseline methods, demonstrating that our theoretical guarantees hold in practice and that our algorithm exhibits strong \textbf{robustness} and \textbf{adaptability} across diverse routing environments.
\end{enumerate}

%% file: related.tex
\subsection{Background and Formal Problem Setup}\label{sec:background}
\noindent\textbf{LLM Routing.}
Existing LLM routing research primarily falls into two paradigms.
The first improves response quality while controlling cost, often through ensembling outputs from different LLMs~\cite{jiang2023llm, wang2023fusing} or cascading strategies that query LLMs sequentially by capability~\cite{chen2023frugalgpt, aggarwal2024automix, yue2023large, lee2023orchestrallm}. These methods incur high latency and cost due to multiple model calls per query.
The second employs learned model-based predictors to estimate the performance or cost for each query and select the optimal LLM~\cite{ong2025routellmlearningroutellms, somerstep2025carrot, hu2024routerbench, ding2024hybrid, lu2023routing, jitkrittum2025universal, shnitzer2023large, stripelis2024tensoropera, feng2024graphrouter, hari2023tryage, wang2025mixllm}, but these approaches introduce nontrivial training overhead. 
Furthermore, adapting them to varying LLM deployment configurations requires retraining, making them unsuitable for dynamic and resource-constrained online routing.
Although recent efforts~\cite{vsakota2024fly, mohammadshahi2024routoo, rama2025cerebrum} formulate LLM routing as MILP, they struggle in high-volume online settings, where queries arrive sequentially, rather than simultaneously, making the offline optimal infeasible to compute.
We complement these works by introducing the first training-free and efficient online routing method with provable performance guarantees, tailored for high-volume, budget-constrained, and dynamic LLM-serving.
See Appendix~\ref{app:related} for extended related work.

%% file: problem.tex

\noindent\textbf{Problem Definition.}
Consider an LLM-serving system deployed with $M$ types of LLMs. 
Each LLM type, indexed by $i \in [M]= \{1, \dots, M\}$, is allocated a running token budget $B_i$ (i.e., representing the limited token budgets available per time unit), with a total budget equaling $B$. 
Typically, the split of the budget is predetermined before the operation of the system.
During each time unit, the system receives a set of queries from various end users, denoted by $Q$. 
The goal is to design an online routing strategy $x(\cdot)$ that assigns incoming queries in $Q$ to the available LLMs in a way that maximizes the overall quality of responses, under the budget constraints.

The offline version of this problem can be naturally formulated as the following MILP~(Objective \ref{eq:prime}),
\begin{wrapfigure}{r}{0.4\textwidth}
\vspace{-1.7em}  
\begin{minipage}{\linewidth}
\small
\begin{equation}\label{eq:prime}
    \begin{split}
    \max & \sum_{j\in Q}\sum_{i\in [M]} d_{ij} x_{ij}  \\
     \text{s.t.} ~ & \sum_{j}  g_{ij}x_{ij} \leq B_i~ \text{   for all $i$,}\\
        & \sum_i x_{ij} \leq 1 ~ \text{   for all $j$,}\\
        & x_{ij} \in \{0, 1\}
\end{split}    
\end{equation}
\end{minipage}
\vspace{-2em}
\end{wrapfigure}
where $\ops$ denotes the \textit{performance score} of the response from LLM $i$ to query $j$, and $g_{ij}$ represents the \emph{token budget consumption} for LLM $i$ processing query $j$ (representing the cost). 
Specifically, $g_{ij}$ can be further decomposed as $ f_i^I \cdot \text{len}(j) + f_i^O \cdot \text{len}(a_{ij})$, where $f_i^I \in \mathbb{R}^M$ and $f_i^O \in \mathbb{R}^M$ represent fixed costs per token during the prefill and decoding stages for LLM $i$, respectively. 


However, solving Objective \ref{eq:prime} (or a natural relaxation) directly in practical online settings poses several non-trivial challenges: 
\textbf{(i) Inaccessible ground-truth performance and cost:} 
    For any query $j$, the true performance score $\ops$ and cost $g_{ij}$ are unavailable without accessing the actual LLMs.
\textbf{(ii) Sequential query arrival under uncertainty:} 
    In practice, queries arrive sequentially rather than simultaneously and must be routed without knowledge of future queries.
\textbf{(iii) Computational scalability:} 
    High query volume demands routing decisions with low latency and minimal computational overhead.
    However, methods that rely on solving large-scale MILP or executing computationally intensive model predictors for each incoming query may violate these constraints.
\textbf{(iv) Deployment scalability:} 
    LLM deployment configurations, such as $M$ or the underlying LLMs, may vary across different environments. 
    Thus, the algorithm must be adaptive to these variations while minimizing adaptation overhead.

\textbf{Motivated by these challenges, we study the following online routing algorithmic problem:} 
Given a larger set of queries $Q$ arriving in a random sequential order,  and a predefined token budget constraint, can we design an online routing algorithm that still achieves a near-optimal cumulative performance?
Formally, we aim to ensure
     $ \frac{\alg}{\opt} \geq 1- o(1) $,
where we define $\alg := \sum_j \sum_i \ops \hat{x}_{ij}$~\footnote{For simplicity, we use the notation $\sum_j$ to denote $\sum_{j \in Q}$ and $\sum_i$ to denote $\sum_{i \in [M]}$, unless stated otherwise.} denotes the total performance of the algorithm with routing results $\hat{x}_{ij}$, and $\opt := \sum_j \sum_i \ops {x}^*_{ij}$ denotes the offline MILP optimum with the optimal solution ${x}^*_{ij}$.

%% file: algo.tex
\input{tables/algo}
\section{Methodology}
To tackle the challenges, our algorithm estimates the performance scores and costs efficiently using a historical dataset (Section~\ref{sec:est}), and learns a routing strategy from a small subset of observed queries (of size $\epsilon |Q|$) to guide future query routing (Section~\ref{sec:learning}).

\subsection{Efficient Performance and Cost Estimation}\label{sec:est}
Our solution to estimate $d$ and $g$ is to leverage a historical dataset $D$ with a specialized data structure that supports efficient similarity-based retrieval. Specifically, let $D=\{j, a_{j}, d_{j}, g_{j}\}_{j=1}^n$, where $a_j \in \mathbb{R}^M$ represents the response vector generated by the $M$  LLMs for query $j$, and $d_{j}, g_{j} \in \mathbb{R}^M$ denotes the corresponding performance scores and costs.

For any incoming query $j \in Q$, we apply a classic ANNS method over $D$ in the embedding space, such as DiskANN~\cite{jayaram2019diskann} and HNSW~\cite{malkov2018efficient}, to select the most similar data points for estimation.
%
This yields a set of approximate nearest neighbors for $j$, denoted by $R_j \subset D$.
The estimated performance score $\eps$ and cost $\hat{g}_{ij}$ for LLM $i$ are then computed as the mean over these neighbors: $\hat{d}_{ij} = \frac{1}{|R_j|} \sum_{q \in R_j} d_{iq}$,~ $\hat{g}_{ij} = \frac{1}{|R_j|} \sum_{q\in R_j} g_{iq}$.

One significant advantage of using ANNS compared with other training-based approaches is the ease of updating the historical dataset $D$ to adopt various deployment configurations without requiring model retraining. This is particularly advantageous for LLM-serving providers, as it incurs almost no additional deployment overhead. 
Online systems can naturally collect and record diverse user queries, which can be directly used to maintain and expand the dataset $D$.
Although collecting ground-truth performance scores and costs typically requires human or automated evaluation, many providers already integrate such mechanisms (e.g., quality feedback from users) in their systems~\citep{chiang2024chatbot}. 
Furthermore, widely used ANNS algorithms in practice, such as HNSW\footnote{We note that many other choices are interchangeable here, e.g. see \url{https://ann-benchmarks.com}.}, use a graph-based indexing structure, making it substantially more efficient in search complexity (typical $O(\log |D|)$ in practice) compared to traditional instance-based methods like exact K-Nearest Neighbor (KNN, $O(|D|)$)~\citep{hu2024routerbench}, achieving significant speedup in search time.

\subsection{Online Routing from Observed Queries}\label{sec:learning}
With the approximate features $\eps$ and $\hat{g}_{ij}$, we propose an online routing algorithm in~\Cref{alg:learn}.

\noindent\textbf{Approximate LP with Control Parameter.}
The first step is to approximate the original MILP using the estimated features $\eps$ and $\hat{g}_{ij}$. 
In addition, we introduce a control parameter $\alpha > 0$ into the objective, leading to the formulation in~\Cref{eq:app_lp}. The inclusion of $\alpha$ does not affect the optimal solution structure -- it simply scales the objective value.
In fact, $\alpha$ acts as a control parameter to ensure generalizability, which is discussed in~\Cref{sec:thm}.

\begin{minipage}{0.32\textwidth}
\small
\begin{equation}
    \begin{aligned}
      \max\ & \sum_{j \in Q} \sum_{i \in [M]}\textcolor{red}{\alpha} \hat{d}_{ij}  x_{ij} \\
        \text{s.t. } 
        & \sum_{j} \hat{g}_{ij} x_{ij} \leq B_i,~ \forall i, \\
        & \sum_{i} x_{ij} \leq 1,~ \forall j, \\
        & x_{ij} \in \{0, 1\},~ \forall i, j
    \end{aligned}
    \label{eq:app_lp}
\end{equation}
\end{minipage}
\hfill
\begin{minipage}{0.32\textwidth}
\small
\begin{equation}
    \begin{aligned}
      \max\ & \sum_{j \in Q} \sum_{i \in [M]}\textcolor{red}{\alpha} \hat{d}_{ij}  x_{ij} \\
        \text{s.t. } 
        & \sum_{j} \hat{g}_{ij} x_{ij} \leq B_i,~ \forall i, \\
        & \sum_{i} x_{ij} \leq 1,~ \forall j, \\
        & \textcolor{red}{x_{ij} \in [0, 1]},~ \forall i, j
    \end{aligned}
    \label{eq:relax}
\end{equation}
\end{minipage}
\hfill
\begin{minipage}{0.34\textwidth}
\small
\begin{equation}
    \begin{aligned}
    \min\ & \sum_{i\in [M]} \gamma_iB_i + \sum_{j\in Q} \beta_j \\
    \text{s.t. } 
    & \beta_j \geq  \textcolor{red}{\alpha} \hat{d}_{ij} - \hat{g}_{ij}\gamma_i, ~\forall i,j,\\
    & \gamma_i \geq 0,\quad \beta_j \geq 0,~ \forall i, j
    \end{aligned}
    \label{eq:dual}
\end{equation}
\end{minipage}

\noindent\textbf{Dual LP under Relaxation.}
Let $s_{max}$ denote the max performance score across all queries, 
and let $\eopt$ be the \emph{offline approximate optimum} obtained by solving~\Cref{eq:app_lp} with $\alpha$ removed.
When the ratio $\eopt/s_{max}$ (or $\opt/s_{max}$) is sufficiently large as $|Q|$ grows, which commonly holds in practice, particularly in high-volume settings, the optimal value to the relaxed LP closely approximates that of the original MILP.~\footnote{In our experiments,  the optimality gap between the relaxed LP and the MILP is only 0.016\% on SPROUT.} This is formalized in Lemma~\ref{lemma:stab} and~\ref{lemma:err} with further discussion in~\ref{app:assump}.
Based on this observation, we apply LP relaxation to~\Cref{eq:app_lp}, allowing $x_{ij}$ to take fractional values in $[0, 1]$.
This yields the relaxed approximate LP in~\Cref{eq:relax} with its dual given in~\Cref{eq:dual}.

\noindent\textbf{Routing via Learned $\gamma^*$.}
By complementary slackness, if $x$ is the optimal solution to~\Cref{eq:relax} and ($\gamma$, $\beta$) is optimal solution to its dual, then $x_{ij} >0 \Leftrightarrow \beta_j = \underset{i}{\max}~ (\alpha \hat{d}_{ij} - \hat{g}_{ij} \gamma_i)$. 
This implies that, at optimality, the dual objective can be expressed as a function parameterized by $\gamma$: ${F}(\gamma) = \sum_{i} \gamma_iB_i + \sum_{j} \underset{i}{\max}~ (\alpha \hat{d}_{ij} - \hat{g}_{ij} \gamma_i)~\refstepcounter{equation}(\theequation)\label{eq:func}$.

Therefore, ideally, given the offline optimal solution $\gamma$, each query $j$ should be routed to the LLM $i$ that maximizes $(\alpha \hat{d}_{ij} - \hat{g}_{ij} \gamma_i)$.
While it is infeasible to compute such a global offline solution in an online setting, this formalization motivates us to treat $\gamma$ as a set of routing weights applied across all LLMs to assist the routing process.
Building on this insight, we adapt this idea to the online setting: we learn an estimated set of weights $\gamma^*$ from the first $P = \epsilon |Q|$ queries. 
The procedure in~\Cref{alg:learn} reflects this idea exactly where its first stage is to learn an optimal $\gamma^*$ that minimizes dual objective ${F}(\gamma, P)$ over the observed queries:  ${F}(\gamma, P) = \epsilon\sum_{i} \gamma_iB_i + \sum_{j\in P} \underset{i}{\max}~ (\alpha\eps - \gamma_i \hat{g}_{ij}) ~\refstepcounter{equation}(\theequation)\label{eq:func_partial}$. 

For each query $j\in P$, the estimated features $\eps$ and $\hat{g}_{ij}$ for all models will be calculated and recorded to solve for $\gamma^*$.
Since this stage requires only a small fraction of the total queries ($\epsilon \ll 1$), random routing is adopted, which may leave some queries in the waiting queue (lines 3, 5-6 in~\Cref{alg:learn}) to improve the generalizability without degrading overall performance.
This approach aligns precisely with the core idea of PAC learning~\cite{denis1998pac, devanur2009adwords}, where uniform and unbiased routing ensures the learned weights $\gamma^*$ generalize to the subsequent routing stage. 
We formalize this in Lemma~\ref{lemma:err} and~\ref{lemma:bound}.
Let the remaining queries be $Y = Q\setminus{P}$.
In the second stage, the learned weights $\gamma^*$ are used to route each incoming query $j \in Y$.
Specifically, each query is directly assigned to the LLM that maximizes the score $(\alpha \eps - \gamma_i^* \hat{g}_{ij})$ (line 12 in~\Cref{alg:learn}). 
If the selected LLM has exhausted its budget, the query is placed in a queue to await execution. 

Compared to existing methods~\cite{ong2025routellmlearningroutellms, somerstep2025carrot, hu2024routerbench}, which often involve repeated heavy computations, our algorithm is significantly more efficient.
It performs optimization only once over a small subset of queries, making it scalable for deployment in high-volume settings.

%% file: tables/algo.tex
\begin{wrapfigure}[12]{R}{0.45\textwidth}
\vspace{-2em}
\begin{minipage}{\linewidth}
\begin{algorithm}[H]
\caption{Routing with Learned $\gamma^*$}
\label{alg:learn}
\small
\begin{algorithmic}[1]
\STATE $P \leftarrow \epsilon Q$ {\small{\texttt{(First $\epsilon$-frac.\ of queries)}}}
\FOR{$j \in P$}
    \STATE Randomly pick $w_j \in \{0\}\cup[M]$
    \STATE Estimate $\hat{d}_{ij}$ and $\hat{g}_{ij}$,  $\forall~i\in [M]$
    \IF{$w_j > 0$}
        \STATE Route $j$ to $w_j$-th LLM
    \ENDIF
\ENDFOR
\STATE Compute $\gamma^* \leftarrow \arg\min_{\gamma} {F}(\gamma, P)$
\STATE $Y \leftarrow Q \setminus P$
\FOR{$j \in Y$}
        \STATE Estimate $\hat{d}_{ij}$ and $\hat{g}_{ij}$,  $\forall~i\in [M]$
        \STATE Compute $\textcolor{red}{\alpha} \hat{d}_{ij} - \hat{g}_{ij}\gamma^*_i$,  $\forall~i\in [M]$
    \STATE Route $j$ to $i = \arg\max_i (\textcolor{red}{\alpha}\hat{d}_{ij} -  \hat{g}_{ij}\gamma^*_i)$
\ENDFOR
\end{algorithmic}
\end{algorithm}
\end{minipage}
\vspace{-1em}
\end{wrapfigure}

%% file: proof.tex
\section{Theoretical Guarantees}\label{sec:thm}
In the following main theorem, we show that our algorithm achieves a competitive ratio close to $1$ compared to the offline opt $\opt$, under natural assumptions (see \Cref{app:assump} for discussion).
\begin{theorem}\label{theorem:main}
    For any given query set $Q$ with random arrival order, \Cref{alg:learn} satisfies
$\frac{\alg}{\opt} \geq 1 - O(\epsilon + \delta)$
   assuming
    $ \frac{\opt}{s_{max}} \geq \Omega(\frac{\alpha M\log(M|\Phi|/\epsilon)}{\epsilon^3(1+\delta)})$, 
where $s_{max}$ is the maximum performance score obtained for any query, and $\Phi$ is an $\epsilon$-net defined over all possible routing strategies $x(\gamma)$. 
\end{theorem}
To establish this theorem, we introduce the following necessary but mild assumption (see~\Cref{app:assump} for a detailed discussion) and auxiliary lemmas.
The key intuition behind approximating features of one query $j$ using a different query $j^\prime$ is that the error between their feature values remains bounded, as long as their embedding representations are close.
We formalize this as follows:
\begin{assumption}\label{assump:dis}
For any query $j,j^\prime$, there exists a value $\eta > 0$ such that if  $||\textsc{Emb}(j)- \textsc{Emb}(j^\prime)||_2 \leq \eta$, then $\forall~i\in [M]$, it holds that $(1- O(\delta))d_{ij^\prime}  \leq \ops \leq (1 + O(\delta)) d_{ij^\prime}$ and $ (1- O(\delta))g_{ij^\prime} \leq {g}_{ij} \leq (1 + O(\delta) ) g_{ij^\prime}$, where $\textsc{Emb}(\cdot)$ represents the embedding function of an embedding model.
\end{assumption}
This approximation strategy naturally introduces estimation errors.
Thus, it is necessary to quantify the discrepancy between the offline optimum $\opt$ and the offline approximate optimum $\eopt$. 
We assume that for any query $j \in Q$, there exists a corresponding feasible set $R_j$ for estimation.
This yields the following Lemma~\ref{lemma:stab}~(proof in~\Cref{app:stab}).
\begin{lemma}\label{lemma:stab}
    Suppose that $~\forall j\in Q$, and $\forall j^\prime \in R_j$, we have $||\textsc{Emb}(j)- \textsc{Emb}(j^\prime)||_2 \leq \eta$, and that~\Cref{assump:dis} holds. Then, the offline approximate optimum $\eopt$ satisfies $\left|\frac{\eopt-\opt}{\opt}\right| \leq O(\delta)$.
\end{lemma}

\Cref{alg:learn} yields the a set of routing results $\hat{x}$, from which we define the \emph{estimated cumulative performance score} as $\est:= \sum_j \sum_i \alpha\eps \hat{x}_{ij}$. 
The results $\hat{x}$ are determined by the learned weights $\gamma^*$ and the budget limitation, and can be represented by $x(\gamma^*)$. 
Accordingly, $\est$ can be expressed as $\est = \sum_i \min\{E_i, \sum_j \alpha \hat{d}_{ij}x_{ij}(\gamma^*)\}$, where $E_i$ represents the maximum feasible contribution under the budget $B_i$, defined as: $E_i := \sum_j^k \alpha \hat{d}_{ij}x_{ij}(\gamma^*)$ with $k = \underset{k}{\argmax} \sum_j^k \hat{g}_{ij} x_{ij}(\gamma^*)$, $~s.t.$~ $\sum_j^k \hat{g}_{ij} x_{ij}(\gamma^*) \leq B_i$. 
We further define the per-LLM estimated performance score $C_{est,i} := \min\{E_i, \sum_j \alpha \hat{d}_{ij}x_{ij}(\gamma^*) \}$.
Analyzing $\est$ thus provides a way to evaluate the routing decisions $x(\gamma^*)$. Unfortunately, directly assessing $\est$ is challenging as it has a complex step function with discontinuities. 
To facilitate analysis, we introduce a relaxed version $\eest := \sum_i \sum_j \alpha \hat{d}_{ij}x_{ij}(\gamma^*)$, which removes the step function present in $\est$.
We further define per-model relaxed estimated performance score $\eest_i := \sum_{j} \alpha \eps x_{ij}(\gamma^*)$, and its counterpart over the observed query set $P$ as $\eest_i(P) := \sum_{j\in P}  \alpha \eps x_{ij}(\gamma^*)$.
Note that for any $j \in P$, we define $x_{ij}(\gamma^*)=1$ if the index $w_j$ randomly selected in the first stage of the algorithm equals $i$; otherwise, $x_{ij}(\gamma^*)=0$.

By removing the discontinuous step function, the analysis becomes tractable.
A key requirement for ensuring the performance of the algorithm is the generalizability of estimated routing weights $\gamma^*$ on the remaining queries $Y$.
To show this, we first prove that the performance disparity between $\eest (P)$ and its expected value $\epsilon\eest$, i.e., $|\eest (P) - \epsilon\eest|$, is bounded.
Since directly bounding this gap over all possible values of $\gamma^*$ is infeasible due to the infinite parameter space, we use an \emph{$\epsilon$-net}~\citep{ xu2024bimetricframeworkfastsimilarity, devanur2009adwords} $\Phi$ (formally defined in \Cref{def:net}) to reduce to a discrete covering set of routing rules, over which we apply a union bound. 
We then extend the analysis to arbitrary $\gamma^*$ values by rounding to $\Phi$. 
The cumulative error over $P$ is bounded in Lemma~\ref{lemma:err} (proof in~\Cref{app:err}):
\begin{lemma}\label{lemma:err}
Let $\Phi$ be an $\epsilon$-net and assume that
    $\frac{\opt}{s_{max}} \geq \Omega(\frac{\alpha M\log(M|\Phi|/\epsilon)}{\epsilon^3 (1+\delta)})$.
If $\forall~ i$, $|\eest_i(P)- \epsilon\eest_i| \leq z_i$,  
then 
$  \sum_i z_i \leq O(\epsilon^2 \sqrt{(1+\delta)\opt\eest/\alpha})$. 
\end{lemma}
The following Lemma demonstrates the generalizability of $\gamma^*$ on remaining queries $Y$ (proof in \Cref{app:mono}). This Lemma highlights the role of the control parameter $\alpha$ in our algorithm: it serves as a bridge to connect the cost and performance.
\begin{lemma}\label{lemma:mono}
If $~\forall i$, it holds that $|\sum_{j\in P} \hat{g}_{ij}x_{ij}(\gamma^*) - \epsilon\sum_j \hat{g}_{ij}x_{ij}(\gamma^*)| \leq O(z_i)$, then there exists a control parameter $\alpha>0$, such that $\forall i$, (1) $|\eest_i(P) - \epsilon\eest_i| \leq z_i$, and (2) $|\eest_i(P) - \epsilon E_i|\leq O(z_i)$.
\end{lemma}
This connection further translates the cost constraint ($B$) to a performance guarantee, which establishes the generalizability of $\gamma^*$ to the remaining queries $Y$, as formalized in Lemma~\ref{lemma:bound}.
\begin{lemma}\label{lemma:bound}
Let $\Phi$ be an $\epsilon$-net. If $\frac{\opt}{s_{max}} \geq \Omega(\frac{\alpha M\log(M|\Phi|/\epsilon)}{\epsilon^3 (1+\delta)})$, then $\est(Y) \geq (1-O(\epsilon)) \est$.
\end{lemma}
Combining the results above, we can prove our main~\Cref{theorem:main}~(proof in~\Cref{app:thm}).

%% file: experiments.tex
\section{Evaluation}\label{sec:eval}
\noindent\textbf{Benchmarks.}
We use 3 different benchmarks in our experiments: RouterBench~\citep{hu2024routerbench}, SPROUT~\citep{somerstep2025carrot}, and Open LLM Leaderboard v2~\citep{fourrier2024open}.
RouterBench contains 11 different LLMs, and we randomly sample 10000 queries as the test query dataset and use the remaining data as the historical dataset.
Open LLM Leaderboard v2 contains 18 different LLMs, where we similarly sample 10000 queries for the test dataset and use the rest as the historical dataset.
SPROUT contains 13 LLMs, and we use the training set as the historical dataset, while the validation and test sets are combined to form the test queries.
In the main setting,  queries are embedded using \texttt{bge-base-en-v1.5}~\cite{bge_embedding}. 
To ensure diversity, we also evaluate with \texttt{SFR-Embedding-2\_R}~\cite{SFR-embedding-2} and \texttt{gte-Qwen2-1.5B-instruct}~\cite{li2023towards}.

\noindent\textbf{Baselines.}
We compare 8 different routing algorithms, classified into two categories: \emph{model-based methods} and \emph{training-free methods}.
For {model-based methods}, following~\citep{somerstep2025carrot}, we train two separate Roberta-based models~\cite{DBLP:journals/corr/abs-1907-11692} to predict generation performance score and cost, yielding:
(i) \textbf{Roberta-perf-routing}, routes each query to the model with the highest predicted performance;
(ii) \textbf{Roberta-cost-routing}, routes each query to the model with the greatest available budget.
{Training-free methods} include:
(iv) \textbf{Random routing};
(v) \textbf{Greedy-perf-routing}, uses ANNS to select the model with the highest predicted performance;
(vi) \textbf{Greedy-cost-routing}, uses ANNS to select the model with the most available budget;
(vii) \textbf{KNN-perf-routing}~\citep{hu2024routerbench}, uses KNN to select the model with the highest performance;
(viii) \textbf{KNN-cost-routing}~\citep{hu2024routerbench}, uses KNN to select the model with the most available budget;
(ix) \textbf{BatchSplit routing}, groups queries into small batches and solves the LP per batch to determine routing. 
We adopt HNSW~\citep{malkov2018efficient} as the main ANNS algorithm and set the number of candidate neighbors ($|R_j|$) to 5 for both ANNS and KNN. 
For BatchSplit, we use a mini-batch size of 256 to balance LP computation cost with the low-latency requirements of online routing.

\noindent\textbf{Metrics.}
Three key metrics are used: 
(1) \emph{Performance}, the total performance score achieved across all test queries;
(2) \emph{Performance per Cost}, the ratio of total performance to cost, reflecting cost efficiency;
(3) \emph{Throughput}, the total number of queries processed, measuring the processing capacity.

\noindent\textbf{Budget.}
We aim to improve online routing under limited budgets in high-volume settings. 
To simulate this, we set the total budget to the minimum cost required for a single model to process all test queries, and vary it by a factor from 0.25 to 2 to evaluate robustness.
We consider several strategies for splitting the total budget across models.
The main setting uses a cost-efficiency-based split, where the total budget is allocated to each model proportionally to its cost efficiency on the historical dataset $D$.
To assess robustness, we also consider uniform, random, extreme, cost-based, and performance-based splits.
For the random split, budgets are randomly assigned and averaged over 100 runs.
In the extreme split, 80\% of the budget is allocated to the $h$ least cost-efficient models ($h = 1$ to $5$), and the remaining 20\% to the others.
For additional experimental setup details, see~\Cref{app:exp}.

\input{tables/main}

\input{figures_tex/queris}
\input{figures_tex/order}

\input{figures_tex/llms}

\input{figures_tex/split}
\noindent\textbf{Main Results.}
\Cref{tab:main} presents the main results under the 3 benchmarks, using $\alpha = 0.0001$ and $\epsilon = 0.025$ for our algorithm, with the maximum available test queries, and historical data in the evaluation.
Our algorithm consistently outperforms all 8 baselines in performance, cost efficiency, and throughput. 
On average, it exceeds all baselines by \textbf{3.55$\times$} in performance, \textbf{1.85$\times$} in cost efficiency, and nearly \textbf{4.25$\times$} in throughput. 
Even against the strongest baseline, BatchSplit, our algorithm still achieves with \textbf{33\%} higher performance, \textbf{38\%} better cost efficiency, and \textbf{24\%} higher throughput across all benchmarks.
We further compare our method to the offline approximate oracle ($\eopt$), and find that it achieves \textbf{75.99\%} to \textbf{84.66\%} of the approximate oracle's performance.
These results align closely with our theoretical guarantees, further validating the effectiveness of our algorithm.

\subsection{Robustness}\label{sec:robust}
\noindent\textbf{Query Volume.}
We vary the number of test queries from 4000 to 12000, serving as different traffic volumes. 
Figure~\ref{fig:queries} clearly shows that our method consistently outperforms all baselines across all benchmarks and metrics.
It scales gracefully, maintaining the highest performance, cost efficiency, and throughput as the query volume increases from low to extremely high loads.
Notably, across all three benchmarks, the performance gap between our method and the strongest baseline (BatchSplit) widens with increasing load, reaching about 50\% higher performance on RouterBench at maximum volume.
These results highlight the strong robustness of our approach under varying query volumes.

\noindent\textbf{Query Arrival Order.}
We evaluate the robustness of our algorithm under varying query arrival orders.
Specifically, we independently shuffle test queries 100 times to simulate realistic, unpredictable online environments.
As shown in Figure~\ref{fig:order_random}, our method consistently outperforms all baselines across metrics and benchmarks under these random permutations, demonstrating strong robustness. 
We further evaluate a worst-case adversarial setting, where expensive queries arrive first with results provided in~\Cref{app:queryorder}, showing that our algorithm still maintains strong performance.

\noindent\textbf{Scalability to LLM Deployments.}
We assess robustness under varying deployment configurations by randomly selecting 2 to 16 LLMs from the Open LLM Leaderboard v2 and repeating each experiment 10 times for diversity.
As shown in Figure~\ref{fig:llm_main}, our algorithm consistently achieves strong performance across configurations, demonstrating strong robustness and adaptability to diverse LLM deployments. 
Additional results on RouterBench and SPROUT (see~\Cref{app:mnum}) further confirm its scalability.

\input{figures_tex/budget}

\begin{figure}[t]
    \centering
    \includegraphics[width=\linewidth]{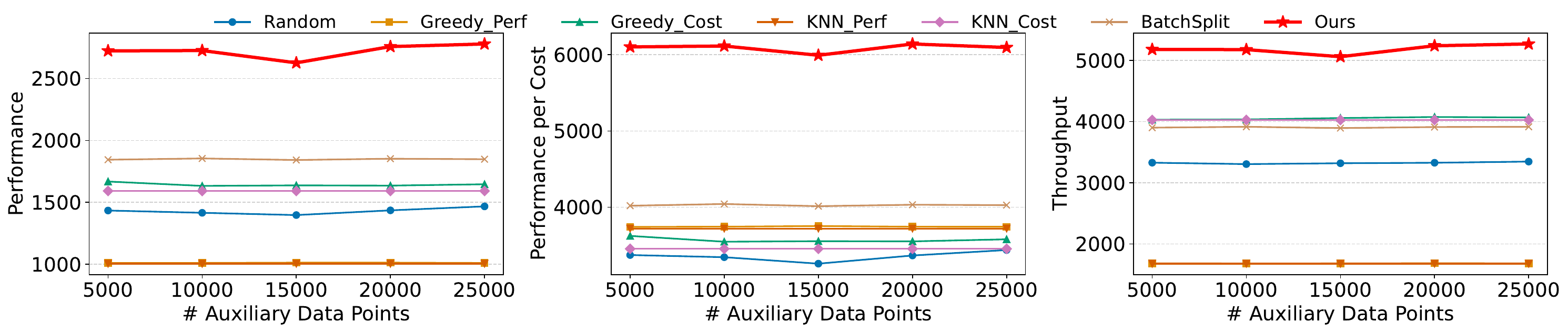}
    \captionsetup{belowskip=-11pt}
    \caption{Results on RouterBench when varying the number of historical data points. 
    }
    \label{fig:basesize_main}
\end{figure}

\begin{figure}[t]
    \centering
    \includegraphics[width=0.98\linewidth]{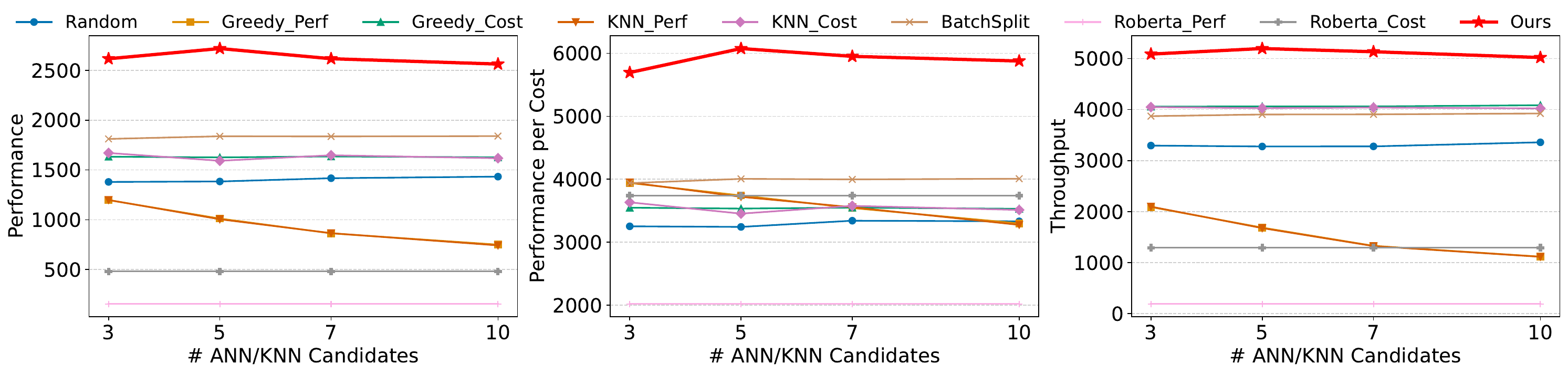}
    \captionsetup{belowskip=-11pt}
    \caption{Results on RouterBench when varying the number of search candidates. 
    }
    \label{fig:topk_main}
\end{figure}

\noindent\textbf{Budget Split.}
We extend the default cost-efficiency-based split to five alternative strategies.
As shown in Figure~\ref{fig:split}, our algorithm consistently outperforms all baselines across all benchmarks and metrics under cost-based, performance-based, uniform, and random splits.
Even under the extreme split scenario (Figure~\ref{fig:extreme_main}), it still achieves strong results.
Notably, when 80\% of the budget is allocated to a single model ($h=1$), our algorithm achieves nearly \textbf{2$\times$} the performance of the strongest baseline (BatchSplit) on RouterBench.
Additional results on SPROUT and Open LLM Leaderboard v2~(see \Cref{app:split}) further highlight its robustness across diverse budget allocation schemes.

\noindent\textbf{Total Budget.}
We vary the total budget $B$ from 0.25 to 2$\times$ the cost of the cheapest model. 
Results shown in Figure~\ref{fig:budget_main} demonstrate that our algorithm consistently outperforms all baselines across metrics on RouterBench as the budget increases.
Even under extremely limited budgets, our algorithm can still achieve the best performance.
Additional results on SPROUT and Open LLM Leaderboard v2 (see~\Cref{app:budget}) further highlight its robustness and adaptability across diverse budget constraints.

\noindent\textbf{Historical Data \& Search Candidates.} 
To evaluate the robustness of our algorithm with respect to the size of historical data, we vary the number of historical data points used in ANNS and KNN from 5000 to 25000 on RouterBench.
As shown in Figure~\ref{fig:basesize_main}, the results clearly show that varying the amount of historical data used in ANNS and KNN has minimal impact on performance, with our algorithm consistently and significantly outperforming all baselines across all settings.
To investigate the impact of the number of search candidates of ANNS and KNN on routing performance, we vary the candidate pool size from 3 to 10 on RouterBench.
As shown in Figure~\ref{fig:topk_main}, the results show that performance is marginally affected, with a slight improvement as the candidate pool size increases. 
Even with as few as 3 search candidates, it outperforms all baselines across benchmarks.
Additional results on other benchmarks (see Appendix~\ref{app:more_res}) further show that our algorithm remains robust under varying historical data sizes and different numbers of ANNS/KNN candidates.

\noindent\textbf{Routing Overhead \& Quality of Historical Data.}
We evaluate the routing latency of~\Cref{alg:learn} under varying query volumes. 
As shown in Appendix Table~\ref{tab:latency}, our method incurs negligible decision overhead compared to existing training-free baselines.
To assess robustness to data quality, we further consider a noisy setting with label perturbations and an out-of-distribution setting with distribution shift.
Results in Appendix Table~\ref{tab:noisy-ood} indicate that our algorithm consistently outperforms all baselines in both settings.
See~\Cref{app:latency} for detailed results.

\subsection{Ablation Studies}\label{sec:abl}
\noindent\textbf{(1) Impact of Embedding Models.}
We evaluate the impact of different embedding models used in ANNS and observe that our algorithm consistently outperforms baselines across all benchmarks (see Appendix Figure~\ref{fig:embed}), demonstrating its robustness to embedding choice.
\noindent\textbf{(2) Impact of $\alpha$ and $\epsilon$.}
We evaluate sensitivity to $\alpha$ and $\epsilon$, as shown in Appendix Figure~\ref{fig:alpha_eps}. 
Performance decreases with larger $\alpha$, with the best result at $0.0001$, aligned with Lemma~\ref{lemma:mono}.
For $\epsilon$, performance increases initially, reaches the peak, and then declines with further increase. 
See~\Cref{app:ablation} for details.

%% file: tables/main.tex
\renewcommand{\arraystretch}{1.3}
\begin{table*}[t]
\small
\caption{
The main results on RouterBench, SPROUT, and  Open LLM Leaderboard v2 are under a total budget equal to the cost of the cheapest model, split across models based on cost efficiency. Here, Perf represents the \emph{Performance}, PPC represents \emph{Performance per Cost}, Tput represents \emph{Throughput}, and RP denotes \emph{Relative Performance} compared with offline approximate optimum ($\eopt$).
}
\setlength{\tabcolsep}{1pt}
  \label{tab:main}
  \centering
  \scalebox{0.78}{\begin{tabular}
    {cccccccccccccccc}
    \noalign{\global\arrayrulewidth1pt}\hline\noalign{\global\arrayrulewidth0.4pt}
   \multirow{2.5}{*}{\normalsize \textbf{Algorithm}}& \multicolumn{5}{c}{\normalsize \textbf{RouterBench}} & \multicolumn{5}{c}{\normalsize \textbf{SPROUT}} & \multicolumn{5}{c}{\normalsize \textbf{Open LLM Leaderboard v2}} \\
    \cmidrule(lr){2-6} \cmidrule(lr){7-11} \cmidrule(lr){12-16}
    & Perf & Cost  & \makecell{PPC} & Tput & \makecell{RP} & Perf & Cost  & \makecell{PPC} & Tput & \makecell{RP} & Perf & Cost  & \makecell{PPC} & Tput & \makecell{RP}\\
    \hline
    \makecell{Random}& 
    1384.25 &        0.427 &    3243.25 &     3276 &  43.10\% &  
    2827.6 &      0.72 &    3927.29 &    4742 &  47.61\% &  
    953.0 &  0.741 &    1284.37 &     2877 &  49.89\% \\
    \hline
   \makecell{Greedy-Perf} & 
   1012.1 &         0.27 &    3742.379 &     1687 &  31.52\% &  
   764.9 &       0.406 &     1881.742 &       1083 &  12.88\% & 
   553.0 &  0.499 &    1107.91 &     1189 &  28.95\%
    \\
    \hline
    \makecell{Greedy-Cost}& 
    1626.25 &        0.46 &   3534.46 &    4061 &  50.64\% & 
    3934.7 &     0.849 &    4630.41 &      6789 &  66.25\% & 
    1051.0 &         0.766 &    1371.30 &    3164 &  55.02\%
    \\
    \hline
    \makecell{KNN-Perf}& 
    1005.1 &         0.27 &    3720.58 &     1677 &  31.3\% &  
    769.6 &      0.407 &    1888.46 &    1084 &  12.96\% & 
    556.0 &  0.498 &    1114.29 &    1194 &  29.11\%
    \\
    \hline
    \makecell{KNN-Cost}& 
    1592.05 &     0.46 &   3454.04 &      4027 &  49.58\% &  
    3905.1 &     0.85 &    4593.37 &     6709 &  65.75\% & 
    991.0 &  0.766 &    1293.07 &     3172 &  51.88\%
    \\
    \hline
    \makecell{BatchSplit}& 
    1838.05 &        0.458 &   4005.93 &    3903 &  57.24\% &  
    3975.5 &      0.83 &    4784.49 &     6221 &  66.94\%    & 
    1059.0 &         0.76 &    1392.07 &    3099 &  55.44\%
    \\
    \hline
    \makecell{Roberta-Perf}& 
    154.5 &  0.077 &   2019.00 &    190 &   4.81\% & 
    458.9 &      0.283 &    1621.64 &    536 &   7.73\%&
    153.0 &  0.207 &   738.21 &     283 &   8.01\%
    \\
    \hline
    \makecell{Roberta-Cost}& 
    481.4 &  0.129 &    3738.88 &    1292 &  14.99\%& 
    3996.2 &     0.848 &    4709.22 &     6765 &  67.29\%     &
    1044.0 &         0.766 &    1362.53 &    3173 &  54.66\%
    \\
    \hline
    \cellcolor{gray!30} \textbf{Ours} & 
    \cellcolor{gray!30}\textbf{2718.6} &   \cellcolor{gray!30}  0.447 &  \cellcolor{gray!30}  \textbf{6075.58} &  \cellcolor{gray!30}  \textbf{5195} &\cellcolor{gray!30}\textbf{84.66\%} & 
    \cellcolor{gray!30}\textbf{4513.05} &   \cellcolor{gray!30}  0.815 & \cellcolor{gray!30}  \textbf{5536.74} &  \cellcolor{gray!30}   \textbf{7475} & \cellcolor{gray!30}\textbf{75.99\%} & 
    \cellcolor{gray!30}\textbf{1465.0} &   \cellcolor{gray!30}  0.711 & \cellcolor{gray!30}  \textbf{2060.3} &  \cellcolor{gray!30}   \textbf{3692} & \cellcolor{gray!30}\textbf{76.7\%} 
    \\
    \hline
    \multicolumn{11}{l}{\textit{\textbf{Offline Oracle (Algorithm Upper Bounds Reference})}} \\
    \hline
    \rowcolor{gray!8}
    \parbox[c][2.5em][c]{2cm}{\centering Approx\\Optimum($\eopt$)}& 
    3211.35 &        0.46 &    6975.16 &     6225 &  100\% & 
    5938.99 &      0.85 &    6986.45 &     8781 &  100\% & 
    1910.0 &         0.765 &    2493.66 &     4319 &  100\%
    \\
    \rowcolor{gray!8}
    \hline
     Optimum ($\opt$) & 
    6376.9 &         0.46 &    13865.62 &    6436 &  198.57\% &  
    11934.4 &     0.848 &    14060.34 &    12336 &         200.94\% & 
    4688.0 &         0.763 &    6143.64 &     4688 &  245.44\%
    \\
    \noalign{\global\arrayrulewidth1pt}\hline\noalign{\global\arrayrulewidth0.4pt}
  \end{tabular}
}
\end{table*}

%% file: figures_tex/queris.tex
\begin{figure}[t]
    \centering
    \includegraphics[width=\linewidth]{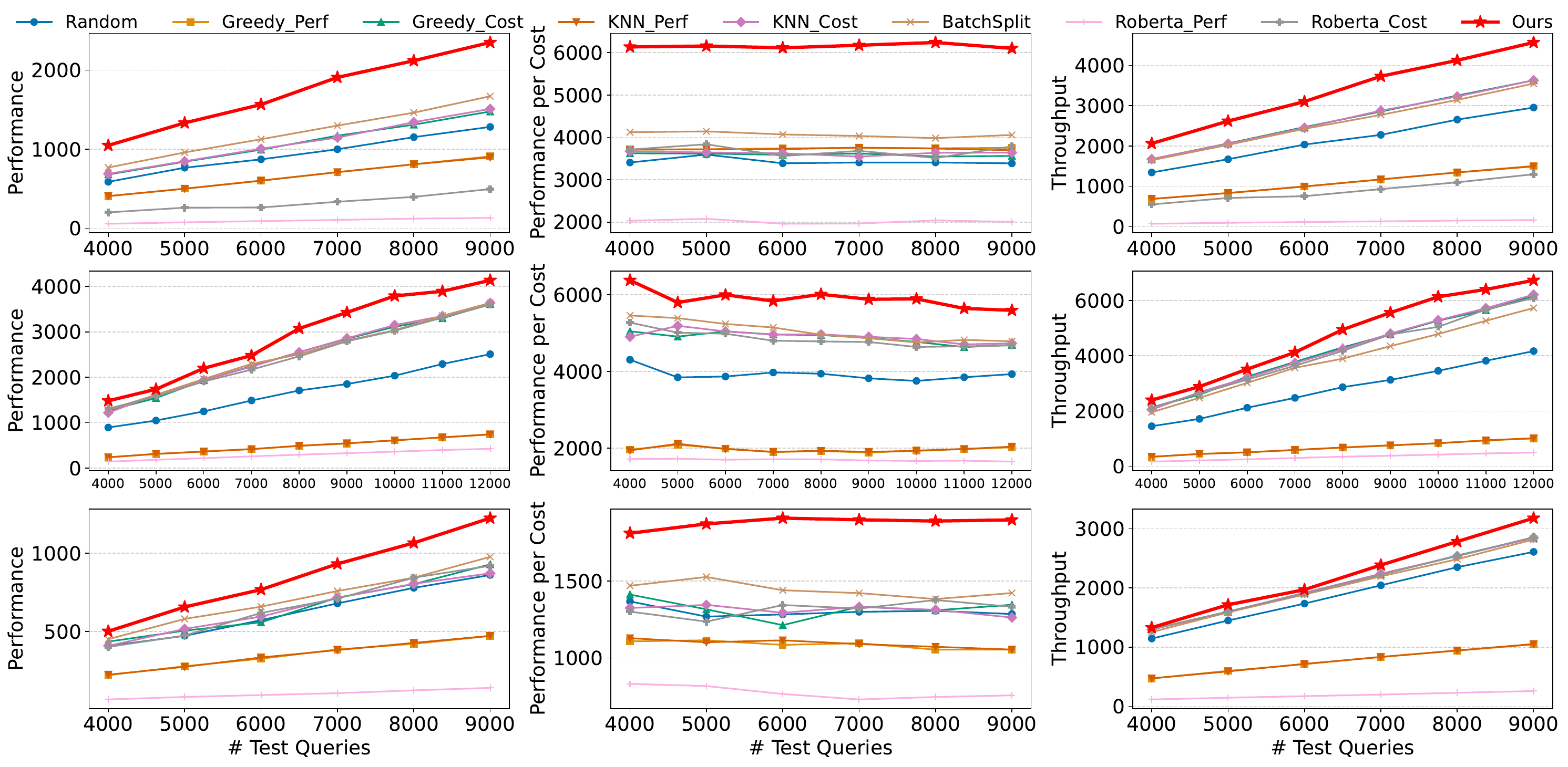}
    \captionsetup{belowskip=-10pt}
    \caption{
    Results with test query volume varying from 4000 to 9000 (12000). Rows correspond to different datasets: RouterBench (top), SPROUT (middle), and Open LLM Leaderboard v2 (bottom).}
    \label{fig:queries}
\end{figure}

%% file: figures_tex/order.tex
\begin{figure}[t]
    \centering
    \includegraphics[width=\linewidth]{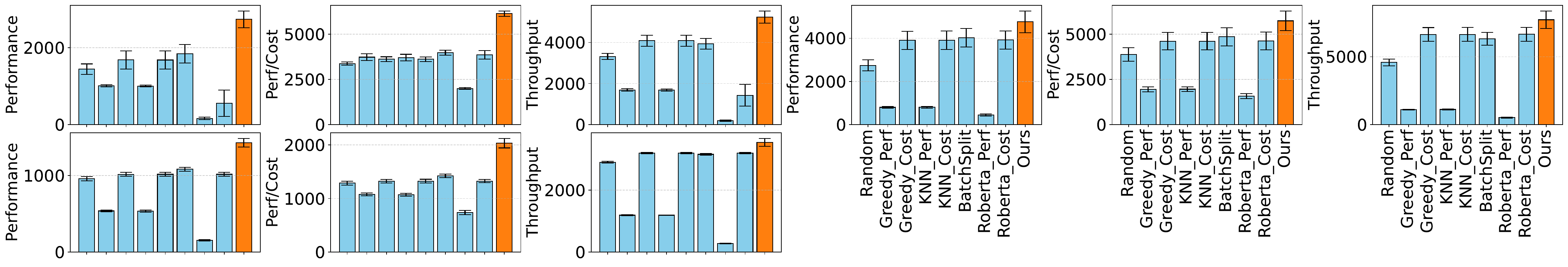}
    \captionsetup{belowskip=-10pt}
    \caption{Results under 100 random query orders.
    (Left to right): the first three subfigures show results on RouterBench, the next three on SPROUT, and the last three on Open LLM Leaderboard v2.}
    \label{fig:order_random}
\end{figure}

%% file: figures_tex/llms.tex
\begin{figure}[t]
    \centering
    \includegraphics[width=\linewidth]{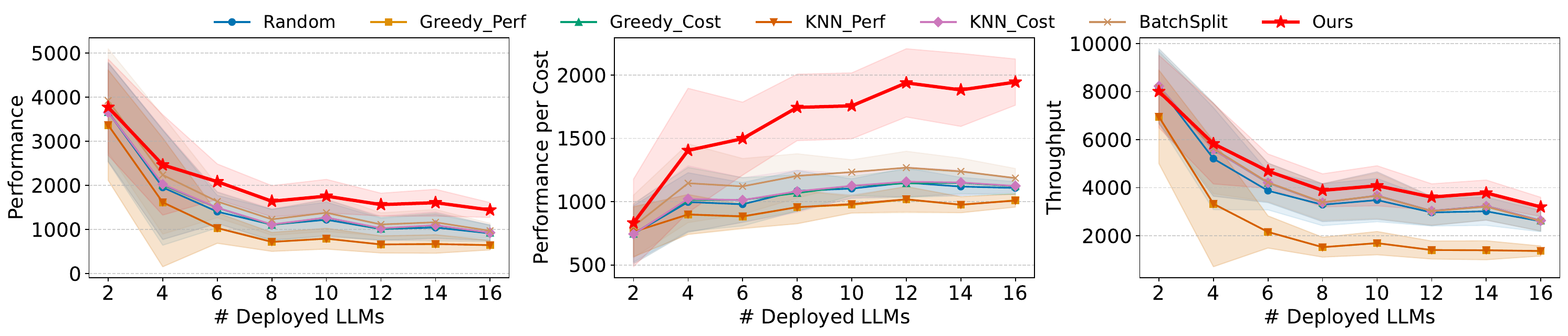}
    \captionsetup{belowskip=-7pt}
    \caption{Results on Open LLM Leaderboard v2 when varying LLM deployment configurations.
    }
    \label{fig:llm_main}
\end{figure}

%% file: figures_tex/split.tex
\begin{figure}[!ht]
    \centering
    \includegraphics[width=\linewidth]{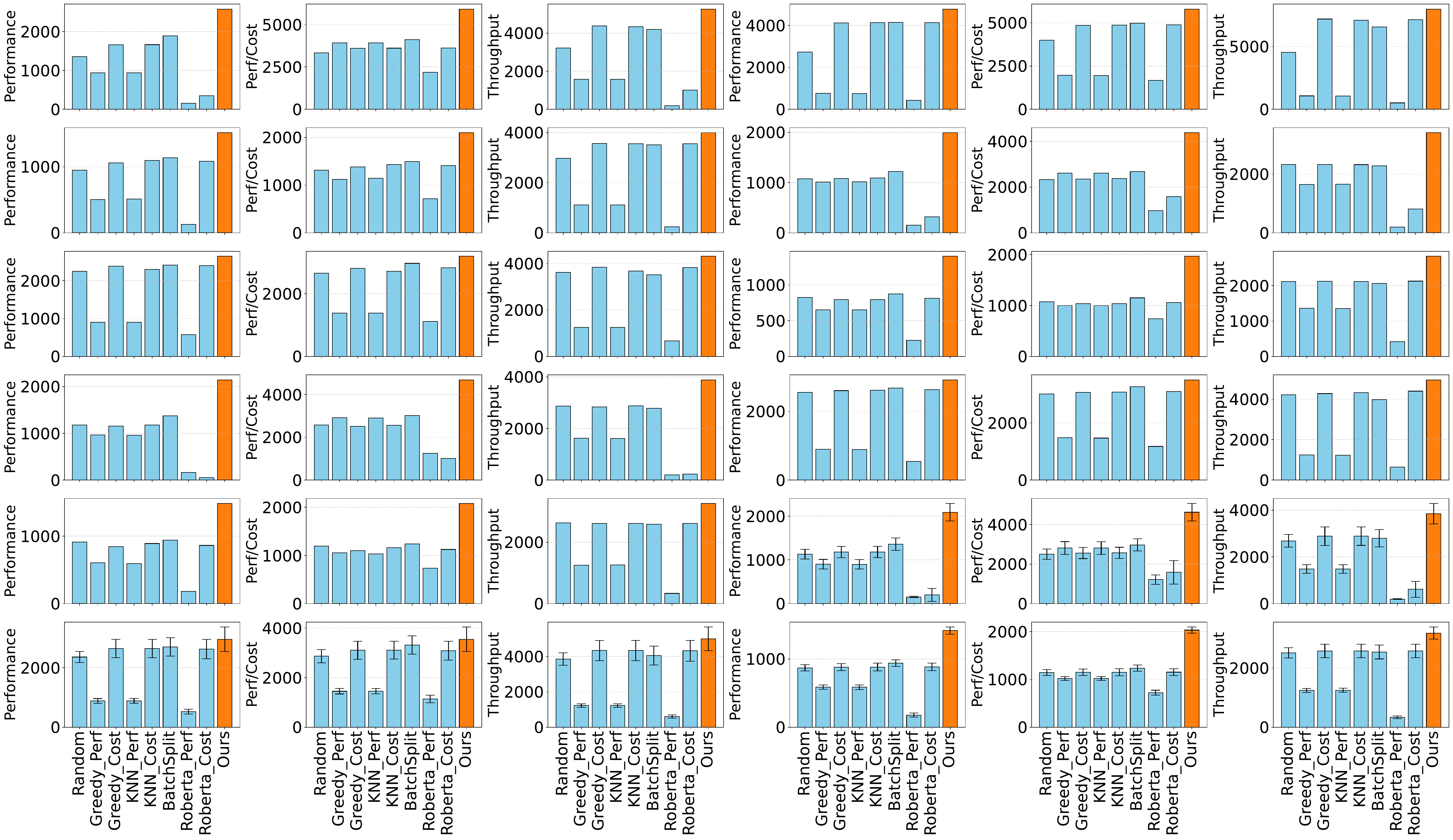}
    \captionsetup{belowskip=-10pt}
    \caption{Results under different budget splitting strategies. 
    Each group of 9 subfigures corresponds to one strategy: 
    (a) cost-based split (subfigures 1–9), 
    (b) performance-based split (10–18), 
    (c) uniform split (19–27), and 
    (d) random split (28–36). 
    Within each group, the first three subfigures correspond to RouterBench, the next three to SPROUT, and the last three to Open LLM Leaderboard v2.}
    \label{fig:split}
\end{figure}

\begin{figure}[t]
    \centering
    \includegraphics[width=\linewidth]{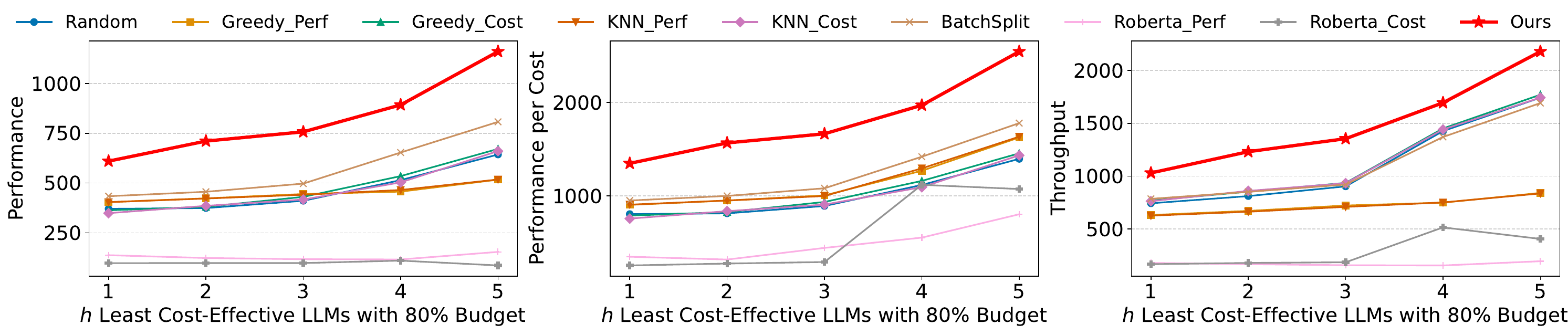}
    \captionsetup{belowskip=-10pt}
    \caption{Results on RouterBench under the extreme budget split.
    }
    \label{fig:extreme_main}
\end{figure}

%% file: figures_tex/budget.tex
\begin{figure}[t]
    \centering
    \includegraphics[width=\linewidth]{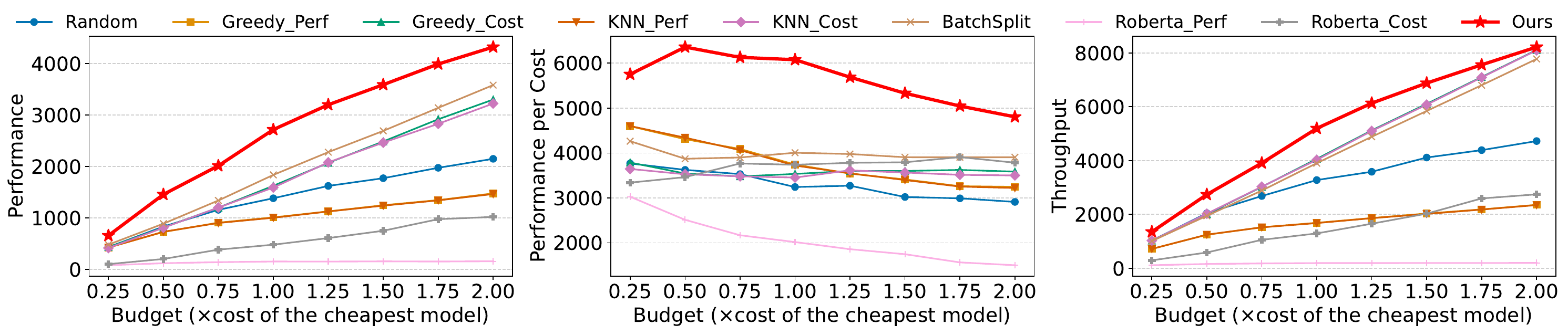}
    \captionsetup{belowskip=-11pt}
    \caption{Results on RouterBench with budget from 0.25 to 2$\times$ the cost of the cheapest model.
    }
    \label{fig:budget_main}
\end{figure}

%% file: discussion.tex
\section{Limitations \& Conclusion}\label{sec:conclusion}
\noindent\textbf{Multi-Factor Routing.} 
In more complex LLM-serving scenarios, additional factors such as query traffic may need to be considered.
However, the primary goal of this work is to develop a cost-efficient online routing algorithm, a problem that has received growing attention in recent literature~\cite{ong2025routellmlearningroutellms, hu2024routerbench, somerstep2025carrot}.
To isolate this core algorithmic problem, we focus solely on two of the most widely studied factors: performance and cost~\cite{ong2025routellmlearningroutellms, hu2024routerbench, somerstep2025carrot}, without explicitly modeling other routing considerations.
Those routing factors can be broadly categorized into two types.
(i) \textbf{Query-intrinsic features}, which are independent of time and query order, such as per-model response quality and cost, can be seamlessly incorporated into our algorithm by adding additional constraints to the original MILP formulation.
Each added constraint introduces a new linear term with a corresponding dual variable in the dual objective, and these dual variables together serve as learnable routing weights for future routing.
(ii) In contrast, \textbf{operational-state features}, which depend on dynamic system conditions (e.g., query traffic or system load), fall outside the scope of our current formulation. 
These factors are more aligned with a related but orthogonal problem: load balancing under traffic constraints.
Nonetheless, we believe our algorithm offers a natural foundation for future extensions that incorporate such traffic-aware adaptations. 
Further discussion is provided in~\Cref{app:dis}.

%% file: conclusion.tex
\noindent\textbf{Conclusion.}
This work presents the first efficient, training-free online routing algorithm for high-volume, budget-constrained LLM serving.
Our algorithm uses ANNS over historical data to efficiently estimate query features and performs a one-time optimization on a small set of observed queries to learn routing weights that ensure strong performance and generalizability, and efficiency.
We theoretically guarantee a competitive ratio of $1 - o(1)$ under mild assumptions, validated by extensive experiments on 3 benchmarks against 8 baselines. 
Our method consistently outperforms all baselines, underscoring strong effectiveness and robustness in diverse online routing scenarios.

%% file: appendix.tex
\section{More Experiment Details}\label{app:exp}
\input{tables/benchmark}

\input{tables/cost}

\noindent\textbf{Benchmarks.}
We evaluate our method on three benchmarks: RouterBench (zero-shot version)~\citep{hu2024routerbench}, SPROUT~\citep{somerstep2025carrot}, and Open LLM Leaderboard v2~\citep{fourrier2024open}.
All three benchmarks are constructed from multiple dataset sources.
Table~\ref{tab:benchmark} summarizes the query types in each benchmark, while Table~\ref{tab:combined_model_costs} details the LLMs used and their corresponding token costs.

For RouterBench, we identify 13 different data sources in a total of 36497 samples -- 6 more than those reported in the original paper -- spanning a diverse range of query domains.
This benchmark includes 11 different LLMs. 
Since it does not provide a predefined train/test split, we randomly sample 10000 queries as the test query set and treat the remaining queries as historical data. 
Note that we do not report per-token costs for models in RouterBench, as~\cite{hu2024routerbench} does not provide this information. Instead, the benchmark includes precomputed costs for each query across all 11 LLMs, which we directly use in our experiments.

For SPROUT, it consists of queries from 6 different datasets, covering different query domains, such as math and RAG. 
It contains a total of 44241 samples across 13 LLMs.
We use the provided training set as the historical dataset, and combine the validation and test sets to construct the test query set.

For Open LLM Leaderboard v2, we follow the same setup and data processing as~\cite{somerstep2025carrot}, resulting in 21065 samples. 
The processed benchmark includes 5 different datasets, with each evaluating different aspects of LLM capabilities.
This benchmark uses 18 different LLMs, and we randomly sample 10000 queries as the test set and use the remainder as the historical data.
Since most evaluations in Open LLM Leaderboard v2 are likelihood-based, the cost is essentially determined by the length of the input.
Following~\cite{somerstep2025carrot}, we report the cost per input token in Table~\ref{tab:combined_model_costs} and compute the total cost of each query based on input token counts.

In the main setting, the queries are embedded using \texttt{bge-base-en-v1.5}~\cite{bge_embedding}. 
For the diversity consideration, we additionally evaluate with two different embedding models: \texttt{SFR-Embedding-2\_R}~\cite{SFR-embedding-2} and \texttt{gte-Qwen2-1.5B-instruct}~\cite{li2023towards}.

\noindent\textbf{Baselines.}
We compare 8 different routing algorithms, categorized into two categories: \emph{model-based methods} and \emph{training-free methods}.

For {Model-based methods}, we follow the training approach in~\citep{somerstep2025carrot} and train two separate Roberta-based models~\cite{DBLP:journals/corr/abs-1907-11692} to predict the performance and cost of each query. 
The performance prediction model is classification-based, aiming to select the optimal LLM with the highest expected performance, while the cost prediction model is regression-based and estimates the expected cost of an incoming query.
This results in 2 different baselines:
\begin{itemize}
\item   \textbf{Roberta-perf-routing}, routes each query to the model with the highest predicted performance;
\item  \textbf{Roberta-cost-routing}, routes each query to the model with the greatest available budget.
\end{itemize}

{Training-free methods} include 6 baselines:
\begin{itemize}
\item  \textbf{Random routing}, randomly selects a model for each query;
\item  \textbf{Greedy-perf-routing}, uses ANNS to select the model with the highest predicted performance;
\item  \textbf{Greedy-cost-routing}, uses ANNS to select the model with the most available budget;
\item  \textbf{KNN-perf-routing}~\citep{hu2024routerbench}, uses KNN to select the model with the highest performance;
\item  \textbf{KNN-cost-routing}~\citep{hu2024routerbench}, uses KNN to select the model with the most available budget;
\item  \textbf{BatchSplit routing}, groups incoming queries into small batches and solves the linear programming (LP) per batch to determine routing. 
\end{itemize}
Note that in the online setting, the true remaining budget of each model is not directly observable during the routing process, as it depends on the actual cost incurred after the query response is generated.
Therefore, all cost-based methods and BatchSplit rely on predicted costs, which are obtained from predictive models, ANNS, or KNN, to estimate the remaining available budget and make routing decisions.

We adopt HNSW~\citep{malkov2018efficient} as the main ANNS algorithm and set the number of candidate neighbors ($|R_j|$) to 5 for both ANNS and KNN.
To assess the robustness of our algorithm under varying candidate set sizes, we further vary this number to 3, 7, and 10.
However, many alternative methods, such as DiskANN~\cite{jayaram2019diskann}, are interchangeable here, as listed in~\url{https://ann-benchmarks.com}.
For BatchSplit, we use a mini-batch size of 256 to balance LP computation cost with the low-latency requirements of the practical online routing.

\noindent\textbf{Metrics.}
In our experiments, we consider three key evaluation metrics:
(1) \emph{Performance}, the overall performance score achieved by processing all test queries under the given budget constraints;
(2) \emph{Performance per Cost}, the ratio of total performance to the corresponding cost, reflecting overall cost efficiency;
(3) \emph{Throughput}, the total number of queries successfully processed, indicating the processing capacity.
We evaluate \emph{Throughput} as a key metric because our goal is to design an effective online routing algorithm for high-volume settings with limited token budgets.
In these scenarios, the available budget may not suffice to serve all incoming queries within a given time unit.
Therefore, \emph{Throughput} reflects how efficiently an algorithm utilizes the available budget to maximize the number of queries successfully served during that current time unit.

\input{tables/perf-cost}
\noindent\textbf{Budget.}
Our algorithm aims to improve routing performance under limited budgets in high-volume settings. 
To simulate this high-volume, budget-constrained setting, we define the total budget based on the minimal cost required for a single model to process all test queries in the benchmark. 
In the main setting, the budget is set to this minimal value. 
To evaluate the robustness of our algorithms, we scale the budget by a factor ranging from 0.25 to 2.

We consider multiple strategies for splitting the total budget across models.
In the main setting, we adopt a cost-efficiency-based split strategy. 
As shown in Table~\ref{tab:routerbench_perf_cost}, \ref{tab:sprout_perf_cost}, and~\ref{tab:leaderboard_perf_cost}, we report the average performance score, cost, and cost efficiency (Perf/Cost) of each model on the historical data across 3 benchmarks.
We observe a substantial disparity in cost efficiency across models.
For instance, for SPROUT, the most efficient model achieves a cost efficiency of 9722, while the least efficient only reaches 108, nearly a 100$\times$ difference. 
Directly allocating the budget based on cost efficiency is thus highly imbalanced, as it can result in nearly all resources being allocated to a single model.
Therefore, we adopt a smoothed version that allocates the budget proportionally to the square root of each model's cost efficiency on the historical data, i.e., according to $(\frac{\textsc{Perf}}{\textsc{Cost}})^{0.5}$. 

In our robustness evaluations, we explore several alternative budget splitting strategies,  including uniform split, random split, extreme split, cost-based split, and performance-based split.
In the cost-based split, the budget is allocated inversely proportional to the average cost.
However, due to the extreme cost imbalance across models (as also observed in the cost-efficiency-based split), we also adopt a smoothed variant that splits the budget proportional to $(\frac{1}{\textsc{Cost}})^{0.5}$.
In the performance-based split, the budget is directly assigned in proportion to the average performance scores of the models.
In the random split, the total budget is randomly allocated across LLMs, and the experiment is repeated 100 times to account for variability.
In the extreme split, 80\% of the budget is assigned to the $h$ least cost-efficient models, and the remaining 20\% is uniformly distributed among the others, where $h$ ranges from 1 to 5.

\noindent\textbf{Optimization Implementation.}
We implement all optimization codes using the CVXPY~\cite{diamond2016cvxpy} package. 
For the one-time optimization step in our algorithm, we use the L-BFGS-B solver~\cite{zhu1997algorithm}. 
For BatchSplit, which involves solving a linear program for each batch, we adopt the HIGHS solver~\cite{huangfu2018parallelizing}.

\noindent\textbf{Devices.}
All experiments are conducted on a machine equipped with 16 CPUs and 32GB of memory. 
For training the Roberta models used in the model-based baselines, we use two NVIDIA H200 GPUs.

\section{Theoretical Proofs}\label{app:proof}

\begin{definition}[\textbf{$\epsilon$-Net}~\citep{devanur2009adwords}]
    Given a parameter $\epsilon > 0$ and routing rule $x(\gamma)$, a set $\Phi \subseteq [0, 1]^M$ is an $\epsilon$-net if, for any $\gamma \in [0,1]^M$, there exists $\gamma^\prime \in \Phi$ such that $\forall~i,j,~ |x_{ij}(\gamma) - x_{ij}(\gamma^\prime)| \leq \epsilon$. \label{def:net}
\end{definition}

\subsection{Discussion on Our Theoretical Assumptions}\label{app:assump}
We introduce several mild assumptions in our theoretical analysis and provide justifications for each below.

\noindent \textbf{Random arrival order.} 
We assume that the queries arrive in a random order, i.e., they can be picked adversarially, but their order is randomly permuted. 
Note that this is a weaker requirement than assuming the queries are sampled i.i.d. from an unknown distribution. This is because if the queries are sampled i.i.d., then they also satisfy the random order requirement, and thus our assumption is weaker. 
Furthermore, the random order model is also a popular assumption in many other online algorithms \cite{mehta2007adwords, devanur2009adwords}, since it typically allows one to go beyond worst-case hardness and obtain meaningful theoretical guarantees. 
In practice, we perform robustness studies by changing the query order and observe that our algorithm still outperforms baselines; see \Cref{sec:robust}.

\noindent\textbf{``$\opt$ is enough large'' is practial.}
In~\Cref{theorem:main}, we assume that $\frac{\opt}{s_{max}} \geq \Omega(\frac{\alpha M\log(M|\Phi|/\epsilon)}{\epsilon^3(1+\delta)})$, which is naturally aligned with high-volume practical settings. 
This is because $\opt$ depends on the total number of queries $|Q|$, whereas our lower bound assumption scales as a function of $M$, the number of LLMs. 
Typically, $M$ can be thought of as a constant while $|Q|$, the total number of queries, should grow over time. 
Indeed, in all of our benchmarks, we observed that for \emph{every query} $j$, the optimal solution $x_{ij}^*$ (that is, the performance score achieved by the optimal solution on query $j$)  satisfies $\sum_{i \in [M]} d_{ij}x_{ij}^* \ge $0.462 consistently on Open LLM Leaderboard v2, 0.639 on RouterBench, and 0.887 on SPROUT, meaning that in practice, $\opt$ scales as $\Omega(Q)$ (linear in $Q$). 
On the other hand, for constant $\epsilon$ and $\alpha$ (e.g. we use $\alpha=0.0001$, $\epsilon =0.025$ in our main setting), our required lower bound is a polynomial of $M$ (more specifically, it is close to a quadratic since $\Phi$ may depend exponentially in $M$ but it is inside a logarithm). 
However, we think of $M$ as a constant. For example, $M \le 18$ in all of our experiments. Even if $M$ is large, it is reasonable to assume that it is orders of magnitude smaller than $|Q|$ (in our experiments, $Q$ is on the order of thousands to tens of thousands, and in real-world systems, it can be much larger).

Our assumption is further corroborated by the relaxation used in~\Cref{sec:est}, where if $\frac{\eopt}{s_{max}}$ is sufficiently large, the original MILP can be closely approximated by its fractional LP relaxation with only a negligible optimality gap.
We validate this in our main experiments over three benchmarks with the observed optimality gaps being: (i) 0.016\% on SPROUT, (ii) 0.086\% on RouterBench, and (iii) 0.3\% on Open LLM Leaderboard v2.

Lastly, we remark that similar reasonable lower-bound assumptions have been made in prior work on the online matching problem~\cite{goel2008online, devanur2009adwords, mehta2007adwords}, where it is commonly assumed that bids for the items are small compared to the total budget.

\noindent\textbf{``Estimating query features via similarity'' is natural.}
The core intuition behind most previous literature on predictive LLM routing~\cite{hu2024routerbench, ong2025routellmlearningroutellms, wang2025mixllm, somerstep2025carrot, stripelis2024tensoropera}, which trains a model-based predictor on auxiliary datasets to estimate performance and cost of future queries, aligns with Assumption~\ref{assump:dis}.
Specifically, these approaches assume that similar queries share similar routing-relevant features.
If no such relationship exists, then historical data would be uninformative, and it would be impossible to leverage historical data for effective future routing. 
In this case, no meaningful inference can be made without directly querying the LLMs, and routing decisions degenerate to random selection.

One may argue that useful routing information can still be obtained by querying only a subset of LLMs.
This idea is echoed in works that adopt cascading strategies, where LLMs are queried sequentially based on their observed capabilities on historical data~\cite{aggarwal2024automix, chen2023frugalgpt}.
However, these approaches also fundamentally rely on a related assumption: LLMs that perform well on historical data or benchmarks are likely to perform well on future queries.
Furthermore, these methods introduce significant latency and computational overhead, as they may consume substantial token budgets before reaching the optimal LLM.
This makes them impractical for high-volume online routing scenarios with tight token budgets and low-latency requirements.

If no useful routing information can be inferred from historical data, no matter whether it is related to model capability or query features, then the system would need to query all LLMs to make informed routing decisions.
This is because in the worst case, under an adversary query targeting the LLM query order in the system, all LLMs may present extremely low performance, except the last LLM. 
In this situation, no strategy can be adopted to improve overall routing performance without assessing all LLMs.
Therefore, it is important and natural to establish our assumption or use a related notion, e.g., Assumption~\ref{assump:dis}, that links historical information with future queries to obtain meaningful routing information and improve routing performance.


\subsection{\texorpdfstring{Proof of Lemma~\ref{lemma:stab}}{Proof of Lemma 2}}\label{app:stab}
\setcounter{theorem}{1} 
\renewcommand{\thelemma}{\arabic{theorem}}
\begin{lemma}
    Suppose that $~\forall j\in Q$, and $\forall j^\prime \in R_j$, we have $||\textsc{Emb}(j)- \textsc{Emb}(j^\prime)||_2 \leq \eta$, and that~\Cref{assump:dis} holds. Then, the offline approximate optimum $\eopt$ satisfies $\left|\frac{\eopt-\opt}{\opt}\right| \leq O(\delta)$.
\end{lemma}

\begin{proof}
Let $x_1$ denote the optimal solution to $\opt$, and $x_2$ the optimal solution to $\eopt$. 

\textbf{Case 1: $\opt\geq \eopt$}. We have 
\begin{align*}
    \left|\frac{\eopt - \opt}{\opt} \right|& = \left|\frac{\sum_{j}\sum_{i} \eps x_{2ij} -  \sum_{j}\sum_{i} \ops x_{1ij}}{\sum_{j}\sum_{i} \ops x_{1ij}}\right|\\
    & \leq  \left|\frac{\sum_j \sum_{i} (\eps - \ops) x_{1ij}}{\sum_{j}\sum_{i} \ops x_{1ij}}\right| \\
    & = \left|\frac{\sum_j \sum_i (\frac{1}{|R_j|}\sum_{q \in R_j} d_{ik} - d_{ij})x_{1ij}}{\sum_{j}\sum_{i} \ops x_{1ij}}\right| \\
    & \leq \left|O(\delta) \frac{\sum_j \sum_i \ops x_{1ij}}{\sum_{j}\sum_{i} \ops x_{1ij}}\right|\\
    & \leq O(\delta)
\end{align*}
where the second inequality follows from the fact that $\sum_{j}\sum_{i} \eps x_{1ij} \leq \sum_{j}\sum_{i} \eps x_{2ij}$ when $x_{2ij}$ is the optimal solution to $\eopt$, and the fourth inequality follows \Cref{assump:dis}.

\textbf{Case 2: $\opt < \eopt$}. We have 
\begin{align*}
    \left|\frac{\eopt - \opt}{\opt} \right|& = \left|\frac{\sum_{j}\sum_{i} \eps x_{2ij} -  \sum_{j}\sum_{i} \ops x_{1ij}}{\sum_{j}\sum_{i} \ops x_{1ij}}\right|\\
    & \leq  \left|\frac{\sum_j \sum_{i} (\eps  - \ops) x_{2ij}}{\sum_{j}\sum_{i} \ops x_{1ij}}\right| \\
    & = \left|\frac{\sum_j \sum_i (\frac{1}{|R_j|}\sum_{q \in R_j} d_{ik} - d_{ij})x_{2ij}}{\sum_{j}\sum_{i} \ops x_{1ij}}\right| \\
    & \leq \left|O(\delta) \frac{\sum_j \sum_i \ops x_{2ij}}{\sum_{j}\sum_{i} \ops x_{1ij}}\right|\\
    & \leq O(\delta)
\end{align*}
where the second inequality uses the fact that $\sum_{j}\sum_{i} \ops x_{2ij} \leq \sum_{j}\sum_{i} \ops x_{1ij}$ when $x_{1ij}$ is the optimal solution to $\opt$, and the fourth inequality follows \Cref{assump:dis} and the final inequality follows $\sum_{j}\sum_{i} \ops x_{2ij} \leq\sum_{j}\sum_{i} \ops x_{1ij}$.
\end{proof}

\subsection{\texorpdfstring{Proof of Lemma~\ref{lemma:err}}{Proof of Lemma 3}}\label{app:err}

\begin{lemma}
Let $\Phi$ be an $\epsilon$-net and assume that
\begin{equation*}
    \frac{\opt}{s_{max}} \geq \Omega(\frac{\alpha M\log(M|\Phi|/\epsilon)}{\epsilon^3 (1+\delta)}).
\end{equation*}
If $\forall~ i$, $|\eest_i(P)- \epsilon\eest_i| \leq z_i$,  
then 
\begin{equation*}
  \sum_i z_i \leq O\left(\epsilon^2 \sqrt{(1+\delta)\opt\eest/\alpha}\right).
\end{equation*}
\end{lemma}
\begin{proof}
Inspired by the techniques used in~\cite{devanur2009adwords, mehta2007adwords, goel2008online}, we first prove that, under the given setting, the partial performance score can be accurately estimated with high probability. 
    Specifically, for each $i \in [M]$ and $\gamma \in \Phi$, we evaluate the following probability  
    \begin{equation*}
        \Pr(|\eest_i(P) - \epsilon\eest_i| > z_{i}).
    \end{equation*}
    Observe that, for any $j \in P$, the variance satisfies \textsc{Var}[$\alpha\eps x_{ij}(\gamma^*)$] $\leq \frac{1}{|Q|}||\eest_i||_2^2$, which implies $\sum_{j\in P}\textsc{Var}[\alpha\eps x_{ij}(\gamma^*)$] $\leq \epsilon ||\eest_i||_2^2$.
    By applying Bernstein's inequality, we obtain 
    \begin{equation*}
        \Pr(|\eest_i(P) - \epsilon\eest_i| > z_i) \leq 2 \exp\left(-\frac{z_i^2/2}{\epsilon||\eest_i||_2^2 + \frac{z_i}{3}\alpha s_{max}}\right).
    \end{equation*}

    We aim to find such $z_i$ that this probability is below a given tolerance $\tau$. 
    Set RHS as $\tau$ and solve for $z_i$, we obtain 
    \begin{equation*}
        z_i = \frac{\frac{2}{3}\log \frac{2}{\tau}\alpha s_{max} + \sqrt{(\frac{2}{3}\log \frac{2}{\tau}\alpha s_{max} + 8\epsilon\log\frac{2}{\tau}||\eest_i||_2^2 } }{2} \leq \frac{2}{3}\log \frac{2}{\tau}\alpha s_{max} + \sqrt{2\epsilon\log\frac{2}{\tau}||\eest_i||_2^2}
    \end{equation*}
    Without affecting the analysis, we define $z_i = \frac{2}{3}\log \frac{2}{\tau}\alpha s_{max} + \sqrt{2\epsilon\log\frac{2}{\tau}||\eest_i||_2^2}$ for simplicity. 
    
    Set $\tau = \epsilon/(M|\Phi|)$, and use $\frac{\opt}{s_{max}} \geq \Omega(\frac{\alpha M\log(M|\Phi|/\epsilon)}{\epsilon^3(1+\delta)})$, we have 
    \begin{align*}
        \sum_i z_i & = \sum_i \frac{2}{3}\log \frac{2}{\tau}\alpha s_{max} + \sum_i\sqrt{2\epsilon\log\frac{2}{\tau}||\eest_i||_2^2}\\
        & \leq O(M\log\frac{1}{\tau}\alpha s_{max}) + \sum_i\sqrt{2\epsilon\log\frac{2}{\tau}\eest_i s_{max}} \\
        & \leq O(\epsilon^3 \opt(1+\delta)) + \sqrt{2M\epsilon\log\frac{2}{\tau}\eest s_{max}} \\
        & \leq O(\epsilon^3 \opt(1+\delta)) +  O(\epsilon^2 \sqrt{(1+\delta)})\sqrt{\opt \eest/\alpha} \\
        & \leq O\left(\epsilon^2 \sqrt{(1+\delta)\opt\eest/\alpha}\right)
    \end{align*}
    This completes the proof.
\end{proof}

\subsection{\texorpdfstring{Proof of Lemma~\ref{lemma:mono}}{Proof of Lemma 4}}\label{app:mono}
\begin{lemma}
If $~\forall i$, it holds that $|\sum_{j\in P} \hat{g}_{ij}x_{ij}(\gamma^*) - \epsilon\sum_j \hat{g}_{ij}x_{ij}(\gamma^*)| \leq O(z_i)$, then there exists a control parameter $\alpha>0$, such that $\forall i$, (1) $|\eest_i(P) - \epsilon\eest_i| \leq z_i$, and (2) $|\eest_i(P) - \epsilon E_i|\leq O(z_i)$.
\end{lemma}
\begin{proof}
    Define $M_i := \sum_{j\in P} \hat{d}_{ij}x_{ij}(\gamma^*) - \epsilon\sum_j \hat{d}_{ij}x_{ij}(\gamma^*)$ and $H_i :=  \sum_{j\in P} \hat{d}_{ij}x_{ij}(\gamma^*) - \epsilon\sum_j^k \hat{d}_{ij}x_{ij}(\gamma^*)$. 
    Then, we aim to ensure the following two conditions hold: (1) $\alpha |M_i| \leq z_i$, and (2) $\alpha|H_i|\leq O(z_i)$.
    
    If $M_i = 0$, condition (1) holds trivially for any $\alpha > 0$; similarly, if $H_i = 0$, condition (2) also holds for any $\alpha > 0$.
    
    Now consider the case where  $M_i \not = 0$, then it requires that $\alpha \leq \frac{z_i}{|M_i|}$, and when $H_i \not = 0$, we need to have $ \alpha \leq \frac{O(z_i)}{|H_i|}$.
    Let $I$ be the set of indice $i$ for which $H_i \not = 0$ or $M_i \not = 0$. For each $i \in I$, we have 
    \begin{equation*}
        V_{i,1} = 
        \begin{cases}
        \frac{ z_i}{|M_i|} & \text{if } M_i \neq 0 \\
        \infty & \text{if } M_i = 0
        \end{cases} 
        \qquad 
        V_{i,2} = 
        \begin{cases}
        \frac{O(z_i)}{|H_i|} & \text{if } H_i \neq 0 \\
        \infty & \text{if } H_i = 0
        \end{cases}
    \end{equation*}
    Let $V_i := \min \{V_{i, 1}, V_{i, 2}\}$, and set 
    \begin{equation*}
        \alpha = \inf_{i\in I} \{V_i\}
    \end{equation*}
    Since $V_i > 0 $ for all $i\in I$, it follows that $\alpha > 0$.
    Therefore, we find a control parameter $\alpha > 0$ such that both conditions (1) and (2) are satisfied.
\end{proof}

\subsection{\texorpdfstring{Proof of Lemma~\ref{lemma:bound}}{Proof of Lemma 5}}\label{app:bound}
\begin{lemma}
Let $\Phi$ be an $\epsilon$-net. If $\frac{\opt}{s_{max}} \geq \Omega(\frac{\alpha M\log(M|\Phi|/\epsilon)}{\epsilon^3 (1+\delta)})$, then $\est(Y) \geq (1-O(\epsilon)) \est$.
\end{lemma}
\begin{proof}
    We first prove that $\forall~i$, it holds that $\max\{\eest_i, F_i(\gamma^*)\} - C_{est, i} \leq \frac{O(z_i)}{\epsilon}$ where 
    \begin{equation*}
        F_i(\gamma^*) := \gamma_i^* B_i + \sum_j (\alpha \eps - \gamma_i^* \hat{g}_{ij})x_{ij}(\gamma^*) 
    \end{equation*}
    From~\Cref{eq:func}, it follows immediately that $\sum_i F_i(\gamma^*) = F(\gamma^*)$. 

    \textbf{Case 1:} $\gamma_i^* > 0$.
    In this case, 
    \begin{equation*}
        \max\{\eest_i, F_i(\gamma^*)\} - C_{est, i} = \max\{\eest_i, \eest_i + \gamma_i^* (B_i - \sum_j\hat{g}_{ij} x_{ij}(\gamma^*)) \} -\min\{E_i, \eest_i\}
    \end{equation*}
    which follows $C_{est, i} = \min\{E_i , \eest_i\}$.
    We consider two sub-cases separately:
    
    \textbf{Sub-case 1.1:} $\sum_j\hat{g}_{ij} x_{ij}(\gamma^*) \leq B_i $. 
    
        Then, 
        \begin{equation*}
            \max\{\eest_i, F_i(\gamma^*)\} - C_{est, i} =  \gamma_i^* (B_i - \sum_j \hat{g}_{ij} x_{ij}(\gamma^*)) \leq B_i - \sum_j \hat{g}_{ij} x_{ij}(\gamma^*)
        \end{equation*}
        Due to complementary slackness conditions over observed queries $P$, we have $\sum_{j\in P} \hat{g}_{ij} x_{ij}(\gamma^*) = \epsilon B_i$. Thus, it follows that 
        \begin{align*}
             & |\sum_{j\in P} \hat{g}_{ij} x_{ij}(\gamma^*) - \epsilon\hat{g}_{ij}x_{ij}(\gamma^*)| = \epsilon | B_i -\hat{g}_{ij}x_{ij}(\gamma^*)| \leq O(z_i) \\
              \Rightarrow & \max\{\eest_i, F_i(\gamma^*)\} - C_{est, i} \leq \frac{O(z_i)}{\epsilon}
        \end{align*}
            
    \textbf{Sub-case 1.2:} $\sum_j\hat{g}_{ij} x_{ij}(\gamma^*) > B_i $. 
    
        We obtain,
        \begin{equation*}
            \max\{\eest_i, F_i(\gamma^*)\} - C_{est, i} =  \eest_i - E_i
        \end{equation*}
        From Lemma~\ref{lemma:mono}, it can be expanded and bounded as:
        \begin{align*}
            & \eest_i - E_i = \frac{1}{\epsilon} (\sum_j \epsilon \alpha \hat{d}_{ij} x_{ij}(\gamma^*) - \epsilon E_i)  \leq  \frac{O(z_i)}{\epsilon} \\
            \Rightarrow & \max\{\eest_i, F_i(\gamma^*)\} - C_{est, i} \leq \frac{O(z_i)}{\epsilon}
        \end{align*}

    \textbf{Case 2:} {$\gamma_i^* = 0$}. Then,
    \begin{align*}
        \max\{\eest_i, F_i(\gamma^*)\} - C_{est, i} = \eest_i - \min\{\eest_i, E_i\} 
    \end{align*}
    When $\sum_j\hat{g}_{ij} x_{ij}(\gamma^*) \leq B_i $, this difference $\eest_i - \min\{\eest_i, E_i\}= 0$.  
    Thus, we only consider the situation $\sum_j\hat{g}_{ij} x_{ij}(\gamma^*) > B_i $, leading to
    \begin{align*}
        \max\{\eest_i, F_i(\gamma^*)\} - C_{est, i} &= \eest_i - \min\{\eest_i, E_i\} \leq \eest_i - E_i \leq \frac{O(z_i)}{\epsilon} 
    \end{align*}
    which follows Lemma~\ref{lemma:mono}.

    \textbf{Lower Bound on} $\est(Y)$.
    From Lemma~\ref{lemma:err} and Lemma~\ref{lemma:mono}, we have, 
    \begin{align*}
        C_{est, i}(Y) & = \min\{\eest_i - \eest_i(P), E_i - \sum_{j\in P} \alpha \hat{d}_{ij}x_{ij}(\gamma^*)  \} \\
        & \geq \min\{(1-\epsilon)\eest_i - z_i,  (1 - \epsilon) E_i - O(z_i)\} \\
        & \geq (1-\epsilon) C_{est, i} - O(z_i)
    \end{align*}
    where second inequality follows that $|\eest_i(Y)- (1-\epsilon)\eest_i| \leq z_i$ and $|\eest_i - E_i| \leq \frac{O(z_i)}{\epsilon}$. 

    Summing over all $i$, we get $\est(Y) \geq (1-\epsilon) \est - \sum_i O(z_i) $. 
    Due to the fact that $\sum_i z_i \leq O\left(\epsilon^2 \sqrt{(1+\delta)\opt\eest/\alpha}\right)$, we finally obtain 
    \begin{align*}
        \est(Y) & \geq (1-\epsilon) \est -O\left(\epsilon^2 \sqrt{(1+\delta)\opt\eest/\alpha}\right)
        \\
        \Rightarrow \est(Y) & \geq (1-O(\epsilon)) \est
    \end{align*}
This completes the proof.
\end{proof}

\subsection{\texorpdfstring{Proof of Theorem~\ref{theorem:main}}{Proof of Theorem 1}}\label{app:thm}
\setcounter{theorem}{0} 
\begin{theorem}
    For any given query set $Q$ with random arrival order, \Cref{alg:learn} satisfies
$\frac{\alg}{\opt} \geq 1 - O(\epsilon + \delta)$
   assuming
    $ \frac{\opt}{s_{max}} \geq \Omega(\frac{\alpha M\log(M|\Phi|/\epsilon)}{\epsilon^3(1+\delta)})$, 
where $s_{max}$ is the maximum performance score obtained for any query, and $\Phi$ is an $\epsilon$-net defined over all possible routing strategies $x(\gamma)$. 
\end{theorem}
\begin{proof}
    
    \textbf{Case $\gamma^* \in \Phi$}. 
    According to Lemma~\ref{lemma:err} and \ref{lemma:bound}, we have a union bound over $\Phi$.
    This union bound implies that with probability greater than $1-\epsilon$, $\sum_i z_i \leq O(\epsilon^2 \sqrt{(1+\delta)\opt\eest})$ holds.
    Therefore, we have that 
    \begin{align*}
        \max \{\eest, F(\gamma^*)\} - \est \leq \frac{1}{\epsilon} \sum_i z_i \Rightarrow  \max \{\eest, F(\gamma^*)\} - \est \leq O\left(\epsilon \sqrt{(1+\delta)\opt\eest/\alpha}\right)
    \end{align*}
    \begin{itemize}
        \item When $(1+\delta)\opt \geq \eest/\alpha$, we have $\max \{\eest, F(\gamma^*)\}- \est  \leq O(\epsilon (1+\delta)\opt)$. 
        
        By weak dualtiy, we have $ \est \leq \alpha \eopt \leq F(\gamma^*)$.
        Also, we have $\frac{\eopt}{\opt} \in [1 - O(\delta), 1 + O(\delta)]$ (from Lemma~\ref{lemma:stab}).
        Therefore, we obtain that $\alpha(1 - O(\delta))\opt - \est \leq O(\epsilon (1+\delta)\opt) \Rightarrow \est/\alpha \geq (1- O(\epsilon+\delta))\opt$.
        \item When $(1+\delta)\opt < \eest/\alpha$, by the same weak dualtiy, we have $\eest - \alpha(1+O(\delta))\opt \leq O(\epsilon\eest/\alpha)$,
        which further implies that $\eest \leq \alpha\frac{1+ O(\delta)}{1 - O(\epsilon)/\alpha} \opt$.
        Therefore, we have $\alpha(1+\delta)\opt - \est \leq O(\epsilon\eest) \Rightarrow \est/\alpha \geq (1- O(\epsilon +\delta))\opt$.
    \end{itemize} 

   We can define $\alg := \sum_i \min\{T_i, \sum_j  {d}_{ij}x_{ij}(\gamma^*)\}$, where $T_i := \sum_j^t  {d}_{ij}x_{ij}(\gamma^*)$ with $t = \underset{t}{\argmax} \sum_j^t {g}_{ij} x_{ij}(\gamma^*)$, $~s.t.$~ $\sum_j^t {g}_{ij} x_{ij}(\gamma^*) \leq B_i$. 
   
   Furthermore, we have $\est/\alpha = \sum_i \min\{{E}_i/\alpha, \sum_j  \hat{d}_{ij}x_{ij}(\gamma^*)\}$, where ${E}_i /\alpha = \sum_j^k  \hat{d}_{ij}x_{ij}(\gamma^*)$ with $k = \underset{k}{\argmax} \sum_j^k \hat{g}_{ij} x_{ij}(\gamma^*)$, $~s.t.$~ $\sum_j^k \hat{g}_{ij} x_{ij}(\gamma^*) \leq B_i$. 

   Let $t_i^*$ be the largest $t$ such that  $\sum_j^{t_i^*} {g}_{ij} x_{ij}(\gamma^*) \leq B_i$.
   According to Assumption~\ref{assump:dis}, for each $i \in [M]$, we have 
   \begin{equation*}
      (1 - O(\delta)) \sum_j^{t_i^*}  \hat{g}_{ij}x_{ij}(\gamma^*)  \leq  \sum_j^{t_i^*} {g}_{ij} x_{ij}(\gamma^*) \leq (1 + O(\delta)) \sum_j^{t_i^*} \hat{g}_{ij}x_{ij}(\gamma^*) 
   \end{equation*}
   Therefore, we have 
   \begin{equation}\label{eq:tail}
       \sum_j^{t_i^*}  \hat{g}_{ij}x_{ij}(\gamma^*) \leq B_i/ (1 - O(\delta)) = (1 + O(\delta)) B_i
   \end{equation}
    Let $k_i^*$ be the largest $k$ such that  $\sum_j^{k_i^*} \hat{g}_{ij} x_{ij}(\gamma^*) \leq B_i$.
    Without loss of generalizability, assume $t_i^* \geq k_i^*$. 
    Then, there exists a set of tail queries, $A_i$, which makes~\Cref{eq:tail} exceed $B_i$ by at most $O(\delta)B_i$.
    Then, we define excess performance contributed by these tail queries as $\sum_{j \in A_i} \hat{d}_{ij}x_{ij}(\gamma^*)$.
    Because the query order is random, we consider the expected excess performance
    \begin{equation*}
        \mathbb{E}[\sum_{j \in A_i} \hat{d}_{ij}x_{ij}(\gamma^*)] = \frac{O(\delta)B_i}{B_i} E_i/\alpha = O(\delta) E_i/\alpha
    \end{equation*}
    and the variance satisfies $\textsc{Var}[\sum_{j \in A_i} \hat{d}_{ij}x_{ij}(\gamma^*)] \leq O(\delta)||E_i/\alpha||_2^2$.
    Using Bernstein's inequality, we have 
    \begin{equation*}
        \Pr(|\sum_{j \in A_i} \hat{d}_{ij}x_{ij}(\gamma^*) -O(\delta) E_i/\alpha| > e_i) \leq 2 \exp\left(-\frac{e_i^2/2}{O(\delta)||E_i/\alpha||_2^2 + \frac{e_i}{3} s_{max}}\right)
    \end{equation*}
    Follow Lemma~\ref{lemma:err}, set RHS as $ \epsilon/(M|\Phi|)$, use $\frac{\opt}{s_{max}} \geq \Omega(\frac{\alpha M\log(M|\Phi|/\epsilon)}{\epsilon^3(1+\delta)})$, and union bound for all $\gamma^*$ and $i$, then with probability greater than $1 - \epsilon$,  
    \begin{equation*}
        \sum_i e_i \leq O\left(\sqrt{\epsilon^3 (1+\delta)O(\delta)\opt E/\alpha^2}\right) 
    \end{equation*}
    where we denote $E = \sum_i E_i$.
    
    Based on this, we have 
    \begin{align*}
        \left|\sum_i T_i - \sum_iE_i/\alpha \right| & \leq \left|\sum_i\sum_j^{t_i^*}  {d}_{ij}x_{ij}(\gamma^*) - \sum_i\sum_j^{k_i^*} \hat{d}_{ij} x_{ij}(\gamma^*) \right| \\
        & \leq \left|(1 + O(\delta)) \sum_i\sum_j^{t_i^*}  \hat{d}_{ij}x_{ij}(\gamma^*) - \sum_i\sum_j^{k_i^*} \hat{d}_{ij} x_{ij}(\gamma^*) \right|\\
        & \leq \left|  O(\delta) \sum_iE_i/\alpha + (1 + O(\delta)) \sum_{j\in A_i} \sum_i\hat{d}_{ij}x_{ij}(\gamma^*)  \right| \\
        & \leq \left|  O(\delta) \sum_iE_i/\alpha + (1 + O(\delta))(O(\delta)E/\alpha + \sum_i e_i)  \right|\\
        & \leq \left(O(\delta) + O\left(\sqrt{\epsilon^3 (1+\delta)O(\delta)\opt/E}\right) \right)  \sum_i E_i /\alpha \\
        & \leq \left(O(\delta) + O\left(\sqrt{\epsilon^3 (1+\delta)O(\delta)/(1 - O(\epsilon + \delta)\alpha) }\right) \right)  \sum_i E_i /\alpha \\
        & \leq \left(O(\delta) + O\left(\epsilon^{3/2} \sqrt{\delta/\alpha} \right) \right)  \sum_i E_i /\alpha 
    \end{align*}
    where the penultimate inequality follows $E/\alpha \geq \est /\alpha \geq (1 - O(\epsilon + \delta))\opt$.
    
    Finally, it leads to $ |C_{alg} - C_{est}/\alpha| \leq  \left(O(\delta) + O\left(\epsilon^{3/2} \sqrt{\delta/\alpha} \right) \right) C_{est}/\alpha$.
    Because $\est/\alpha \geq (1- O(\epsilon +\delta))\opt$, we obtain that $\frac{\alg}{\opt} \geq 1- O(\epsilon +\delta)$.

    \textbf{Case $\gamma^* \not \in \Phi$}. 
    As $\Phi$ is a $\epsilon$-net, there exists an $\gamma^\prime$ such that $\forall~ i,j,~|x_{ij}(\gamma^*) -x_{ij}(\gamma^\prime)| \leq \epsilon $. 
    Therefore, we obtain 
    \begin{align*}
        |\eest_i(\gamma^*, P)- \epsilon\eest_i(\gamma^*)| & \leq |\eest_i(\gamma^\prime, P)- \epsilon\eest_i(\gamma^\prime)| +|\eest_i(\gamma^\prime, P)- \eest_i(\gamma^*, P)| +|\epsilon\eest_i(\gamma^\prime)- \epsilon\eest_i(\gamma^*)| \\
        & \leq z_i^\prime + \epsilon\eest_i(\gamma^*, P) +\epsilon^2\eest_i(\gamma^*) \\
        & \leq z_i^\prime + O(\epsilon^2 \eest_i(\gamma^*))
    \end{align*}
    
    Similarly, for each $i$, we have
    \begin{align*}
        |\sum_{j \in A_i} \hat{d}_{ij}x_{ij}(\gamma^*) -O(\delta) E_i(\gamma^*)/\alpha| & \leq |\sum_{j \in A_i} \hat{d}_{ij}x_{ij}(\gamma^\prime) -O(\delta) E_i(\gamma^\prime)/\alpha| \\ 
        & \qquad +|\sum_{j \in A_i} \hat{d}_{ij}x_{ij}(\gamma^\prime) - \sum_{j \in A_i} \hat{d}_{ij}x_{ij}(\gamma^*)| \\
        & \qquad + |O(\delta) E_i(\gamma^\prime)/\alpha - O(\delta) E_i(\gamma^*)/\alpha| \\
        & \leq e_i^\prime + \epsilon\sum_{j \in A_i} \hat{d}_{ij}x_{ij}(\gamma^*) + \epsilon O(\delta)E_i(\gamma^*)/\alpha \\
        & \leq e_i^\prime + O(\epsilon\delta E_i(\gamma^*)/\alpha)
    \end{align*}

    where, with slight abuse of notation, we use $E_i(\gamma^*)$ to denote the value computed under $\gamma^*$ and $E_i(\gamma')$ under $\gamma'$.

    By summing over all $i$, and following Lemma~\ref{lemma:err} as well as the proof in case $\gamma^*\in \Phi$, we obtain 
    $\est/\alpha  \geq (1- O(\epsilon +\delta))\opt$, and $ |C_{alg} - C_{est}/\alpha| \leq  \left(O(\delta) + O\left(\epsilon^{3/2} \sqrt{\delta/\alpha} \right) \right) C_{est}/\alpha$.
    
    Therefore, \Cref{alg:learn} satisfies that $\frac{\alg}{\opt}\geq 1 - O(\epsilon +\delta)$.
\end{proof}

\section{Additional Experimental Results}\label{app:res}

\begin{figure}[ht]
    \centering
    \includegraphics[width=\linewidth]{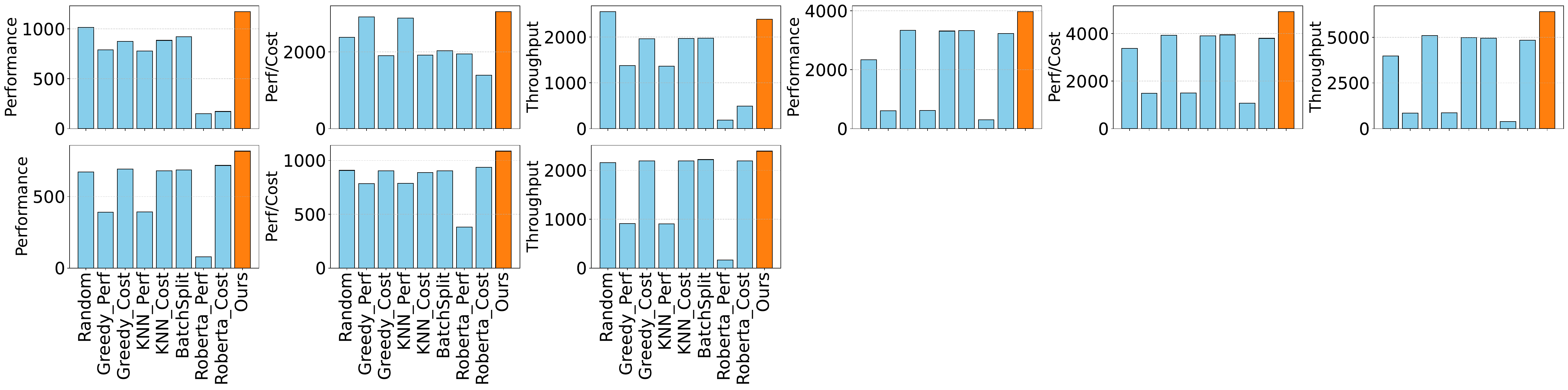}
    \caption{Results under the worst-case query order, where queries are sorted in descending order of cost.
    (Left to right): the first three subfigures correspond to RouterBench, the next three to SPROUT, and the last three to Open LLM Leaderboard v2.}
    \label{fig:order_worst}
\end{figure}

\begin{figure}[ht]
    \centering
    \includegraphics[width=\linewidth]{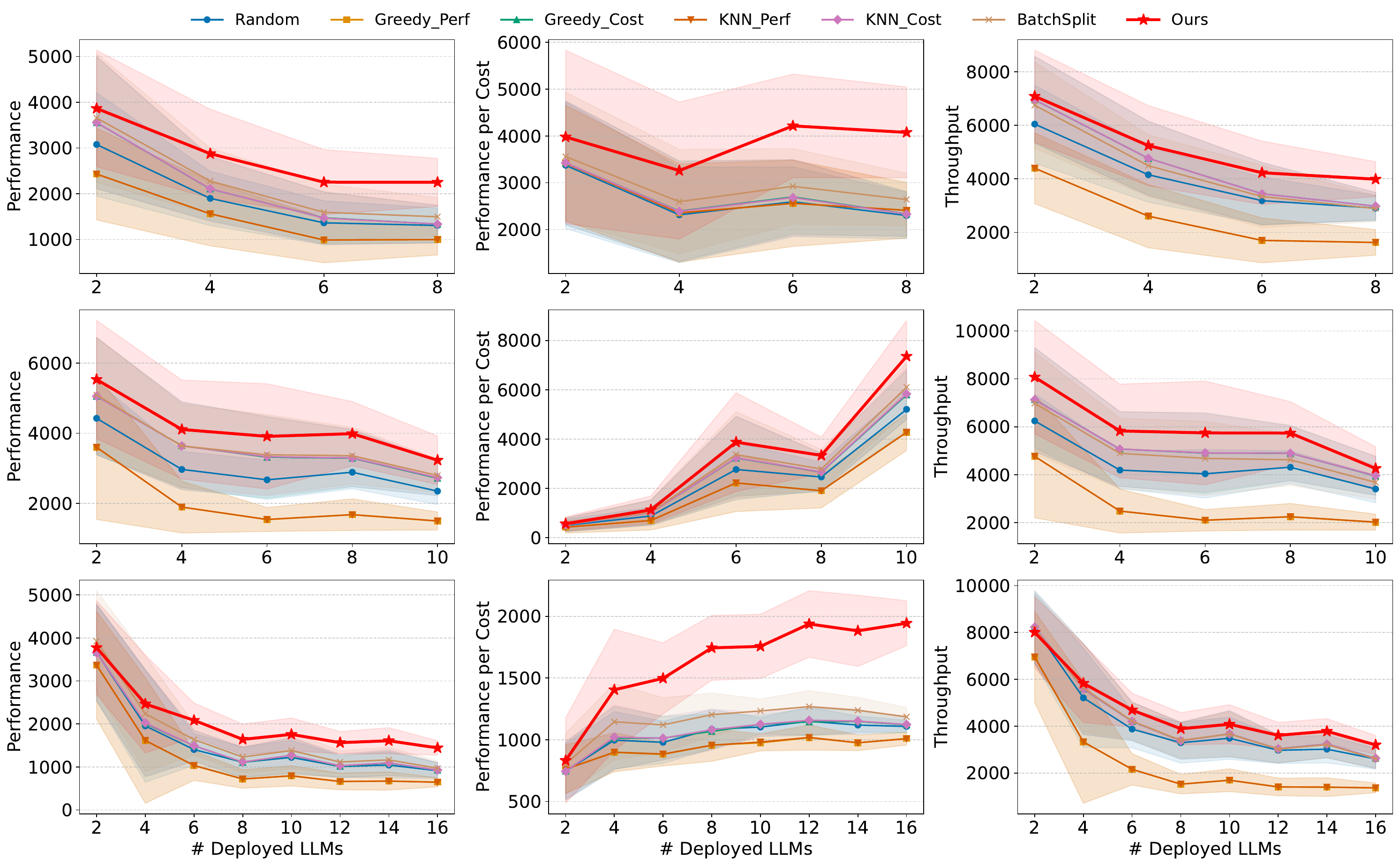}
    \caption{Results when varying the configurations of deployed LLMs.
    Rows correspond to different datasets: RouterBench (top), SPROUT (middle), and Open LLM Leaderboard v2 (bottom). 
    }
    \label{fig:llm}
\end{figure}

\subsection{Query Arrival Order}\label{app:queryorder}
We evaluate the robustness of our algorithm under varying query arrival orders.
Specifically, we consider two settings:
(1) a randomized setting, where test queries are independently shuffled 100 times to reflect realistic, unpredictable online environments; and
(2) a worst-case setting, where queries are sorted in descending order of their maximum cost across all models. 
This adversarial order simulates scenarios in which expensive queries arrive early and are more likely to exhaust the budget.
Figure~\ref{fig:order_random} shows that our algorithm consistently outperforms all baselines across metrics and benchmarks under random permutations.
Even under the worst-case order (Figure~\ref{fig:order_worst}), our method maintains the advantage, achieving the best performance across all metrics on SPROUT and Open LLM Leaderboard v2, and outperforming all baselines in terms of performance and cost-efficiency on RouterBench.
These results demonstrate the robustness of our algorithm towards varying query orders.

\begin{figure}[H]
    \centering
    \includegraphics[width=\linewidth]{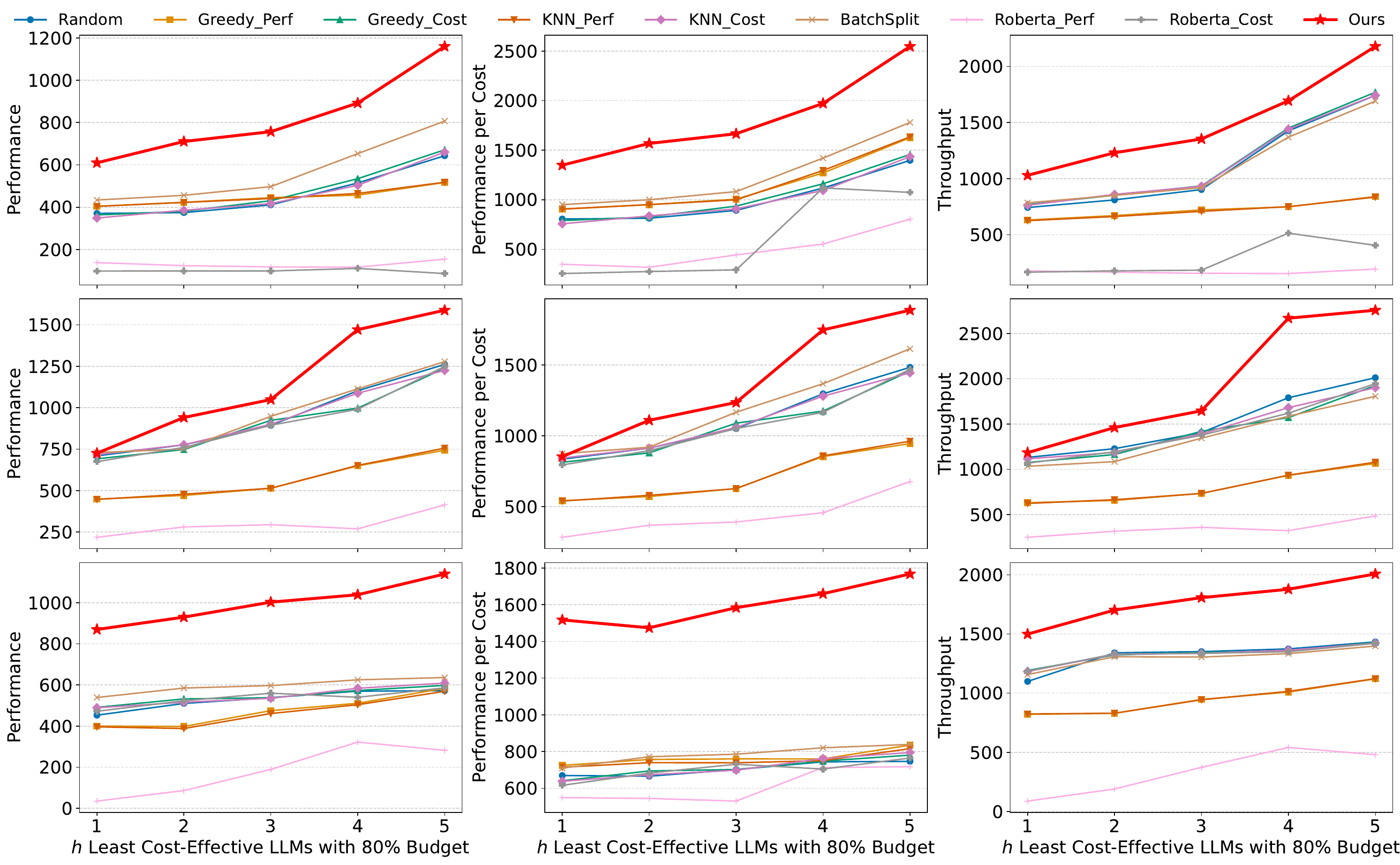}
    \caption{Results under the extreme budget split.
    Rows correspond to different datasets: RouterBench (top), SPROUT (middle), and Open LLM Leaderboard v2 (bottom).
    }
    \label{fig:extreme}
\end{figure}
\begin{figure}[t]
    \centering
    \includegraphics[width=\linewidth]{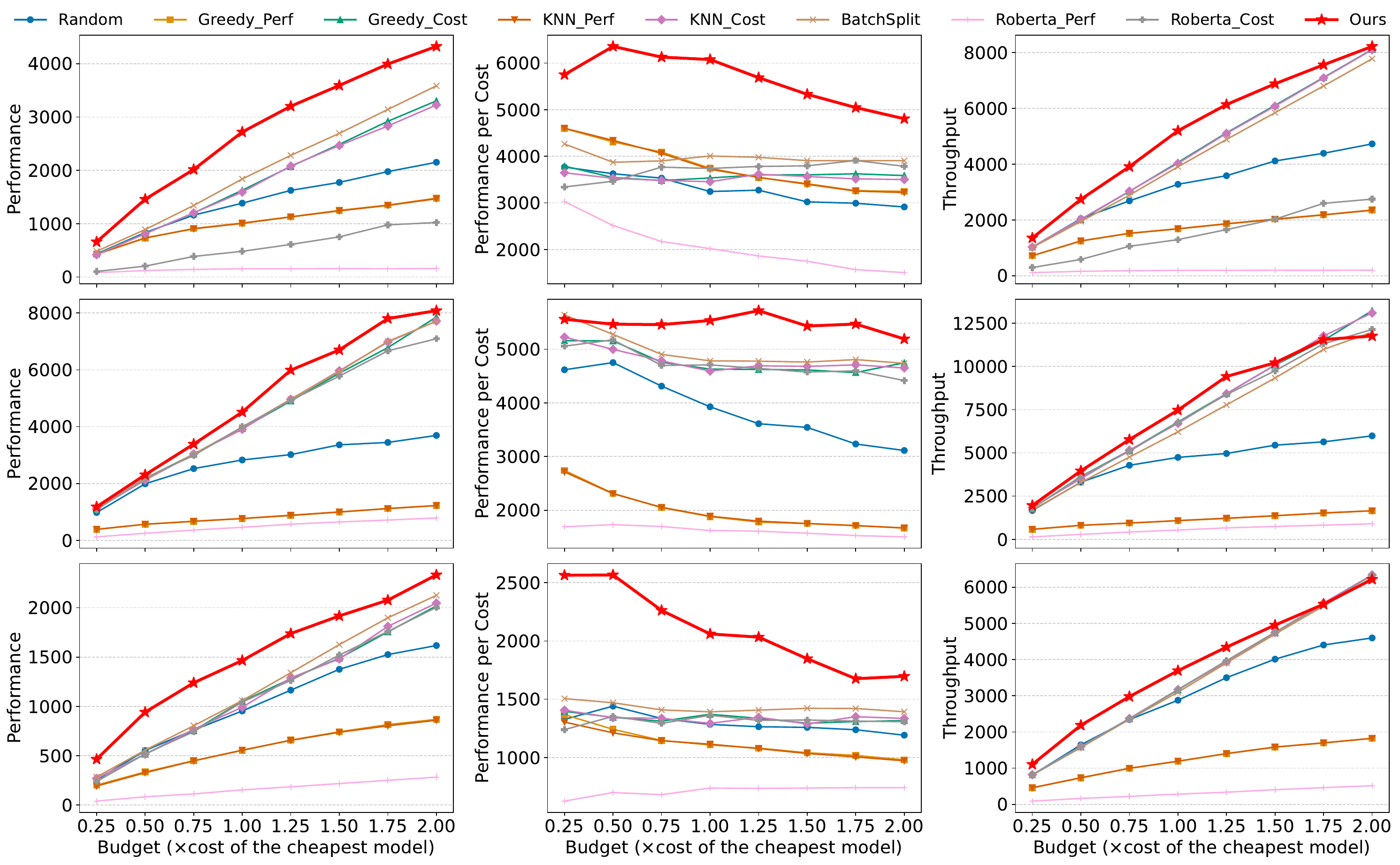}
    \caption{Results when varying total budget $B$ from 0.25 to 2$\times$ the cost of the cheapest model. Rows correspond to different datasets: RouterBench (top), SPROUT (middle), and Open LLM Leaderboard v2 (bottom).
    }
    \label{fig:budget}
\end{figure}

\subsection{Scalability to LLM Deployments.}\label{app:mnum}
One of the key advantages of our algorithm is its scalability to varying configurations of LLM deployments.
To evaluate the robustness of our algorithm, we vary the configurations of deployed LLMs on each benchmark.
For RouterBench, which includes 11 distinct LLMs in total, we vary the deployed LLMs from 2 to 8.
For SPROUT, which has 13 different LLMs, we vary the deployed LLMs from 2 to 10.
For Open LLM Leaderboard v2, which includes 18 distinct models, we vary the number of deployed LLMs from 2 to 16.
In all settings where the number of deployed LLMs is fewer than the maximum, we randomly sample the deployed LLMs and repeat the experiment 10 times to ensure diversity and coverage of possible configurations.
Note that in this experiment, we compare only against training-free methods, as retraining models used in model-based methods for each of the numerous settings would incur significant computational and deployment overhead.
As shown in Figure~\ref{fig:llm}, our algorithm consistently achieves strong performance across all deployment configurations, where the cost-efficiency gap over other baselines steadily increases. 
These results demonstrate the robustness and adaptability of our method to diverse and dynamic LLM serving environments.

\input{figures_tex/basedata}

\input{figures_tex/topk}
\subsection{Budget Split}\label{app:split}
One key factor affecting the performance of routing algorithms is the choice of budget split strategy.
We extend the default cost-efficiency-based split to five alternative strategies.
As shown in Figure~\ref{fig:split}, our algorithm consistently outperforms all baselines across all benchmarks and metrics under four different split strategies: cost-based, performance-based, uniform, and random.
Even under the extreme split setting (Figure~\ref{fig:extreme}), where a large portion of the budget is concentrated on a few low-efficiency models, our method maintains leading performance.
Notably, when 80\% of the budget is allocated to a single model ($h=1$), our algorithm achieves nearly \textbf{2$\times$} the performance of the strongest baseline (BatchSplit) on the RouterBench.
On Open LLM Leaderboard v2, it also demonstrates approximately \textbf{2$\times$} higher cost efficiency than all baselines.
These results underscore the strong robustness and adaptability of our algorithm against a wide range of complex and imbalanced budget allocation schemes.

\input{figures_tex/embed}

\input{figures_tex/alpha}
\subsection{Total Budget}\label{app:budget}
To evaluate the performance of our algorithms in different budget settings, we scale the total budget $B$ from 0.25 to 2$\times$ the cost of the cheapest model. 
Results shown in Figure~\ref{fig:budget} demonstrate that our algorithm achieves competitive or superior performance compared to all baselines across benchmarks and metrics.
Notably, even under the extremely constrained budget, it maintains a clear advantage.
For instance, at $B = 0.25$, it achieves nearly \textbf{2$\times$} the performance and cost efficiency of the strongest baseline (BatchSplit) on the Open LLM Leaderboard v2. 
These results underscore the robustness of our method and its adaptability to different resource availability scenarios.

\subsection{Historical Data Size \& ANNS/KNN Candidates}\label{app:more_res}
\noindent\textbf{Historical Data Size.}
To evaluate the robustness of our algorithm with respect to the size of historical data, we vary the number of historical data points used in ANNS and KNN from 5000 to 25000.
For RouterBench, whose maximum available historical data is 26497, we randomly sample the subset of data points from 5000 to 25000.
For RouterBench, which contains up to 26497 historical records, we randomly sample subsets within this range.
For SPROUT, with a maximum of 30968 records, we similarly sample subsets from 5000 to 25000.
For Open LLM Leaderboard v2, which has 11065 historical records, we sample subsets ranging from 5000 to 10000.
As shown in Figure~\ref{fig:basesize}, the results clearly show that varying the amount of historical data used in ANNS and KNN has minimal impact on performance, with our algorithm consistently and significantly outperforming all baselines across all settings.
This demonstrates the robustness and efficiency of our algorithm: even with as few as 5000 historical records, it remains effective for routing.

\noindent\textbf{Number of ANNS/KNN Candidates.}
To investigate the impact of the number of search candidates of ANNS and KNN on routing performance, we vary the candidate pool size from the default 5 to 3, 7, and 10.
As shown in Figure~\ref{fig:topk}, the results show that performance is marginally affected, with a slight improvement as the candidate pool size increases. 
Notably, our algorithm consistently maintains a leading position across all settings.
Even with as few as 3 search candidates, it outperforms all baselines across benchmarks, demonstrating its stability and robustness to the choice of candidate pool size.

\begin{table}[ht]
\centering
\small
\caption{Per-query routing latency (ms) of different algorithms on RouterBench under varying numbers of queries.}
\label{tab:latency}
\begin{tabular}{lcccc}
\toprule
Algorithm & 4000 queries & 6000 queries & 8000 queries & 10000 queries \\
\midrule
Greedy-Perf   & 0.203 & 0.203 & 0.201 & 0.203 \\
Greedy-Cost   & 0.233 & 0.230 & 0.226 & 0.233 \\
KNN-Perf      & 0.478 & 0.475 & 0.469 & 0.475 \\
KNN-Cost      & 0.300 & 0.282 & 0.281 & 0.278 \\
BatchSplit    & 0.300 & 0.282 & 0.281 & 0.278 \\
\textbf{Ours} & \textbf{0.060} & \textbf{0.060} & \textbf{0.060} & \textbf{0.064} \\
\bottomrule
\end{tabular}
\end{table}

\begin{table}[t]
\centering
\caption{Routing performance on RouterBench under two challenging historical-data conditions: {noisy data label} and {out-of-distribution (OOD) data}.
Perf represents the \emph{Performance}, PPC represents \emph{Performance per Cost}, and Tput represents \emph{Throughput}.
}
\label{tab:noisy-ood}
\small
\begin{tabular}{lcccccccc}
\toprule
& \multicolumn{4}{c}{Noisy Data Label} & \multicolumn{4}{c}{Out-of-Distribution (OOD)} \\
\cmidrule(lr){2-5} \cmidrule(lr){6-9}
Algorithm & Perf & Cost & PPC & Tput & Perf & Cost & PPC & Tput \\
\midrule
Random         & 1403.250 & 0.419 & 3350.430 & 3227   & 1495.35 & 0.518 & 2888.14 & 3308.00 \\
Greedy-Perf    & 1323.600 & 0.370 & 3573.750 & 2471   & 807.15  & 0.252 & 3204.76 & 1643.00 \\
Greedy-Cost    & 1660.200 & 0.461 & 3604.720 & 4047   & 1677.45 & 0.561 & 2989.30 & 4044.00 \\
KNN-Perf       & 1323.950 & 0.369 & 3585.390 & 2472   & 816.10  & 0.254 & 3212.31 & 1658.00 \\
KNN-Cost       & 1660.200 & 0.461 & 3604.720 & 4047   & 1679.70 & 0.561 & 2993.23 & 4047.00 \\
BatchSplit     & 1757.250 & 0.456 & 3856.680 & 3974   & 1828.55 & 0.526 & 3474.42 & 4283.00 \\
Roberta-Perf   & 83.000   & 0.045 & 1836.330 & 84     & 9.50    & 0.017 & 559.15  & 10.00   \\
Roberta-Cost   & 358.000  & 0.075 & 4787.593 & 1616   & 1692.65 & 0.562 & 3013.92 & 4037.00 \\
\textbf{Ours}  & \textbf{2319.650} & 0.440 & \textbf{5221.280} & \textbf{4775} & \textbf{2050.8} & 0.532 & \textbf{3851.53} & \textbf{4644.00} \\
\bottomrule
\end{tabular}
\end{table}

\subsection{Routing Overhead \& Quality of Historical Data}\label{app:latency}
\textbf{Routing Overhead.} 
We evaluate the routing latency of~\Cref{alg:learn} under varying query volumes. 
We compare~\Cref{alg:learn} with training-free routing baselines, focusing on the core online routing stage that follows the embedding step.
This focus is motivated by two key reasons:
(1) The embedding step can operate asynchronously and is fully decoupled from the core routing logic, where its outputs are shareable across retrieval, logging, and other downstream tasks.
(2) The embedding model is interchangeable (e.g., MiniLM, GTE-small, bge-small), while our core contribution lies in the core routing algorithm itself.
As shown in Table~\ref{tab:latency}, our method incurs negligible decision overhead compared to existing training-free baselines.
Even under high query volume (10,000 queries), it introduces only 0.064ms of routing latency per query and consistently achieves the lowest decision overhead (except the random routing strategy, which exhibits significantly worse performance).

\textbf{Quality of Historical Data.}
To assess robustness to data quality, we consider a noisy setting with label perturbations and an out-of-distribution (OOD) setting with distribution shift.
In the noisy setting, we introduce two types of strong noise simultaneously to the data label. 
For each performance label $d_{ij}$, we randomly flip its value with a probability of 20\%. 
For each cost label $g_{ij}$, we apply multiplicative, mean-preserving noise consisting of: (i) log-normal jitter $\left(\log \mathcal{N}(0, \sigma^2) - \frac{1}{2}\sigma^2\right)$ with $\sigma = 0.25$, and (ii) a spike factor of 3.0 applied with 2\% probability.
In the OOD setting, we construct a distribution shift using RouterBench, which contains 13 diverse datasets. 
Specifically, we use MMLU as the historical dataset and evaluate on the remaining 12 datasets, which differ significantly in data distribution.
Results in Table~\ref{tab:noisy-ood} show that our algorithm consistently outperforms all baselines under both noisy label and OOD settings. 
It achieves the highest overall performance, the best performance-per-cost ratio, and the greatest throughput in both settings, demonstrating strong robustness to label noise and distribution shift.

\begin{figure}[t]
    \centering
    \includegraphics[width=0.9\linewidth]{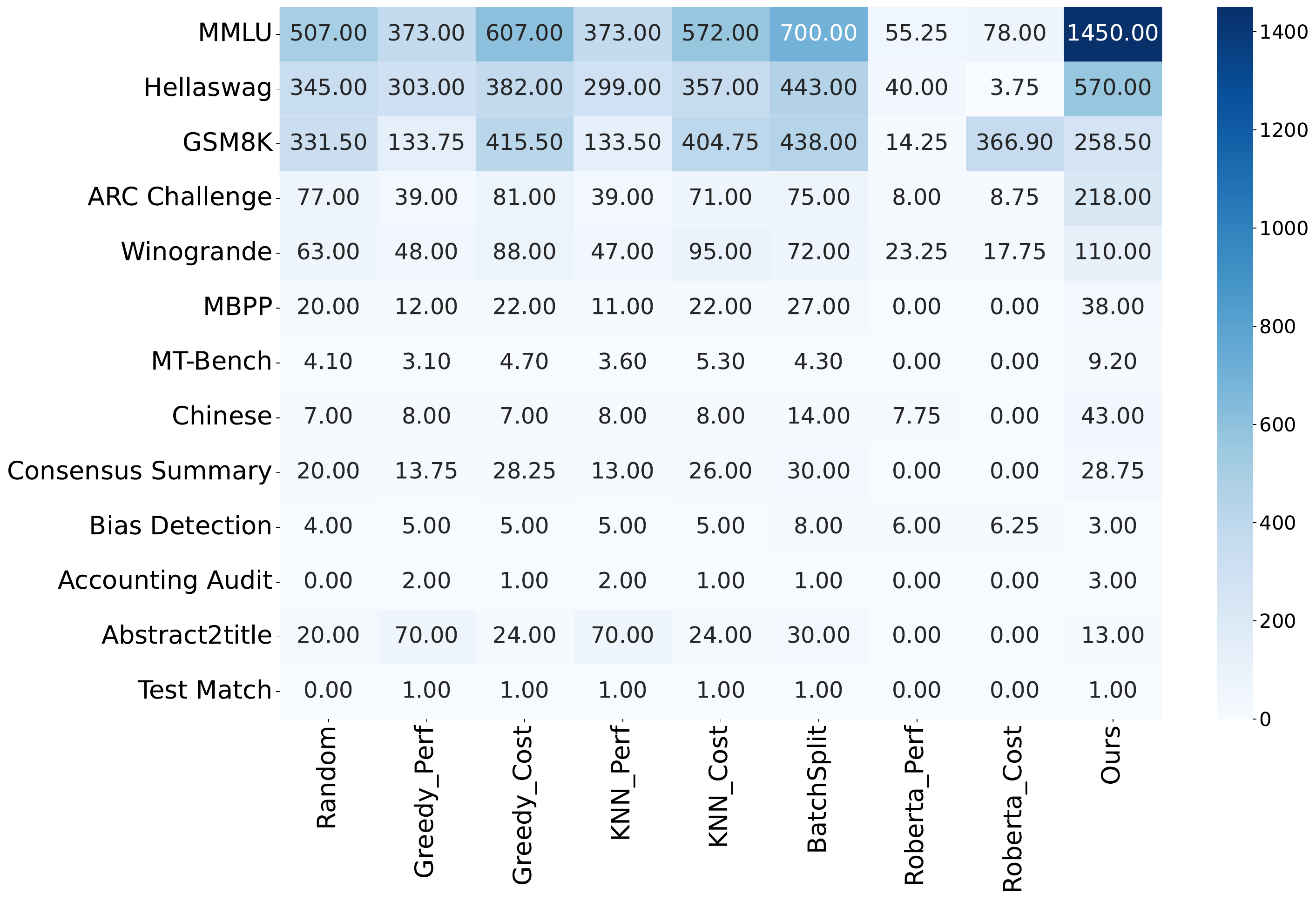}
     \caption{Results across different query types in RouterBench.
    }
    \label{fig:rheat}
\end{figure}

\subsection{Detailed Ablation Study}\label{app:ablation}
\noindent\textbf{Impact of embedding models.}
The choice of embedding model is also important for estimating the performance score and cost of incoming queries. 
To evaluate the adaptivity of our algorithm to different embedding models used in ANNS, we conduct experiments with two larger embedding models: \texttt{SFR-Embedding-2\_R}~\cite{SFR-embedding-2} and \texttt{gte-Qwen2-1.5B-instruct}~\cite{li2023towards}.
As shown in~Figure~\ref{fig:embed}, our algorithm consistently outperforms all baselines across all benchmarks, demonstrating its robustness to the choice of embedding model.

\noindent\textbf{Impact of $\alpha$ and $\epsilon$.}
We evaluate the sensitivity of our method to the parameters $\alpha$ and $\epsilon$, as shown in Figure~\ref{fig:alpha_eps}. 
For $\alpha$, we observe a consistent decline in performance as it increases, with the best result achieved at $\alpha = 0.0001$. 
This trend aligns with Lemma~\ref{lemma:mono}, which shows that a larger $\alpha$ amplifies the gap between the performance of our algorithm on the observed subset $P$, $\eest(P)$, and its expected value $\epsilon \eest$. 
This increased discrepancy results in greater deviation in performance on future queries, ultimately leading to a reduction in overall performance on the full query set $Q$.
For the parameter $\epsilon$, we observe that increasing its value initially improves performance, reaching a peak at around $\epsilon = 0.025$, after which further increases lead to a decline.
This aligns with the intuition that $\epsilon$ controls the number of samples used in the learning stage: too few samples lead to underfitting, while too many may lead to overfitting to the observed data.

\begin{figure}[t]
    \centering
    \includegraphics[width=0.9\linewidth]{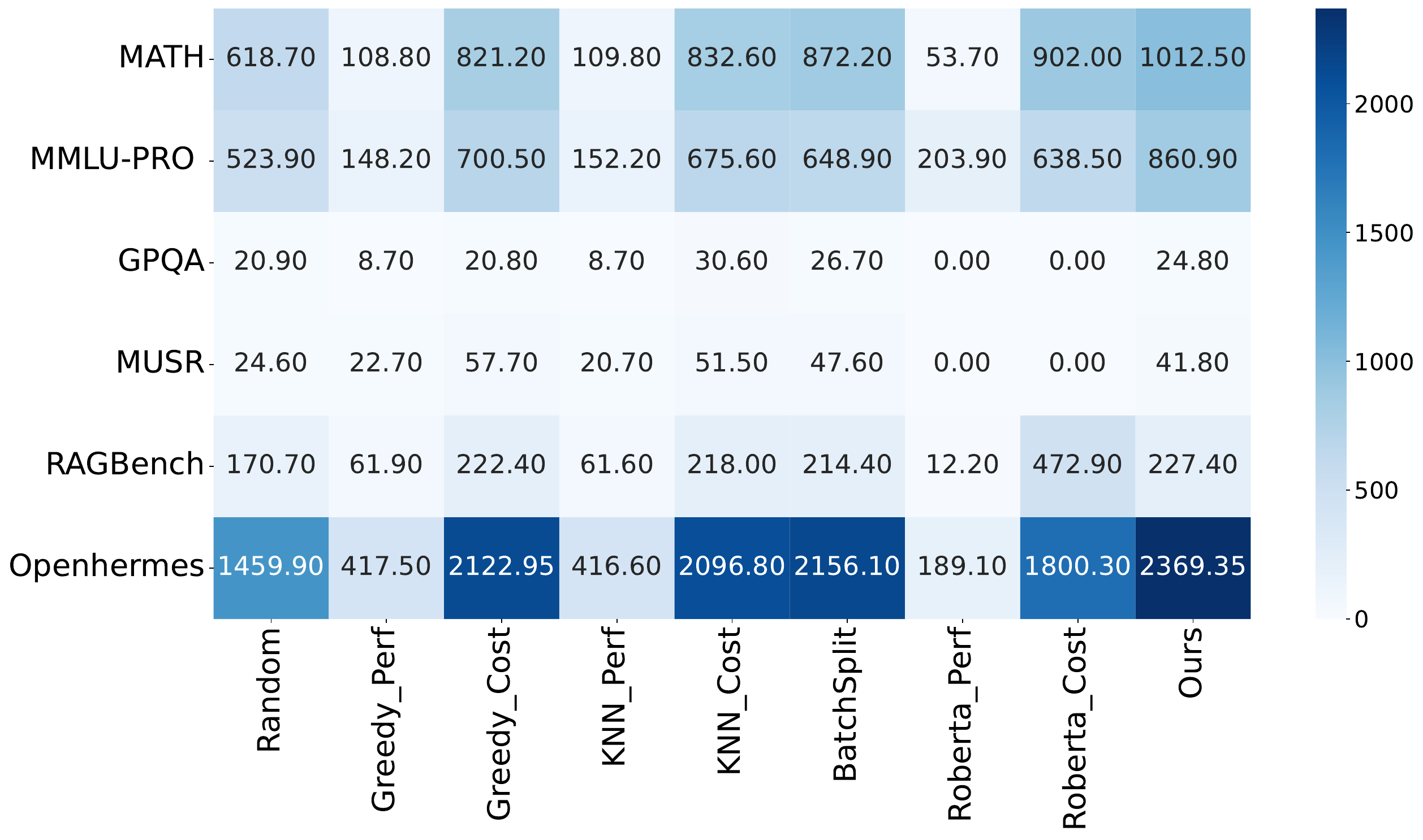}
     \caption{ Results across different query types in SPROUT. 
    }
    \label{fig:sheat}
\end{figure}

\begin{figure}[t]
    \centering
    \includegraphics[width=0.9\linewidth]{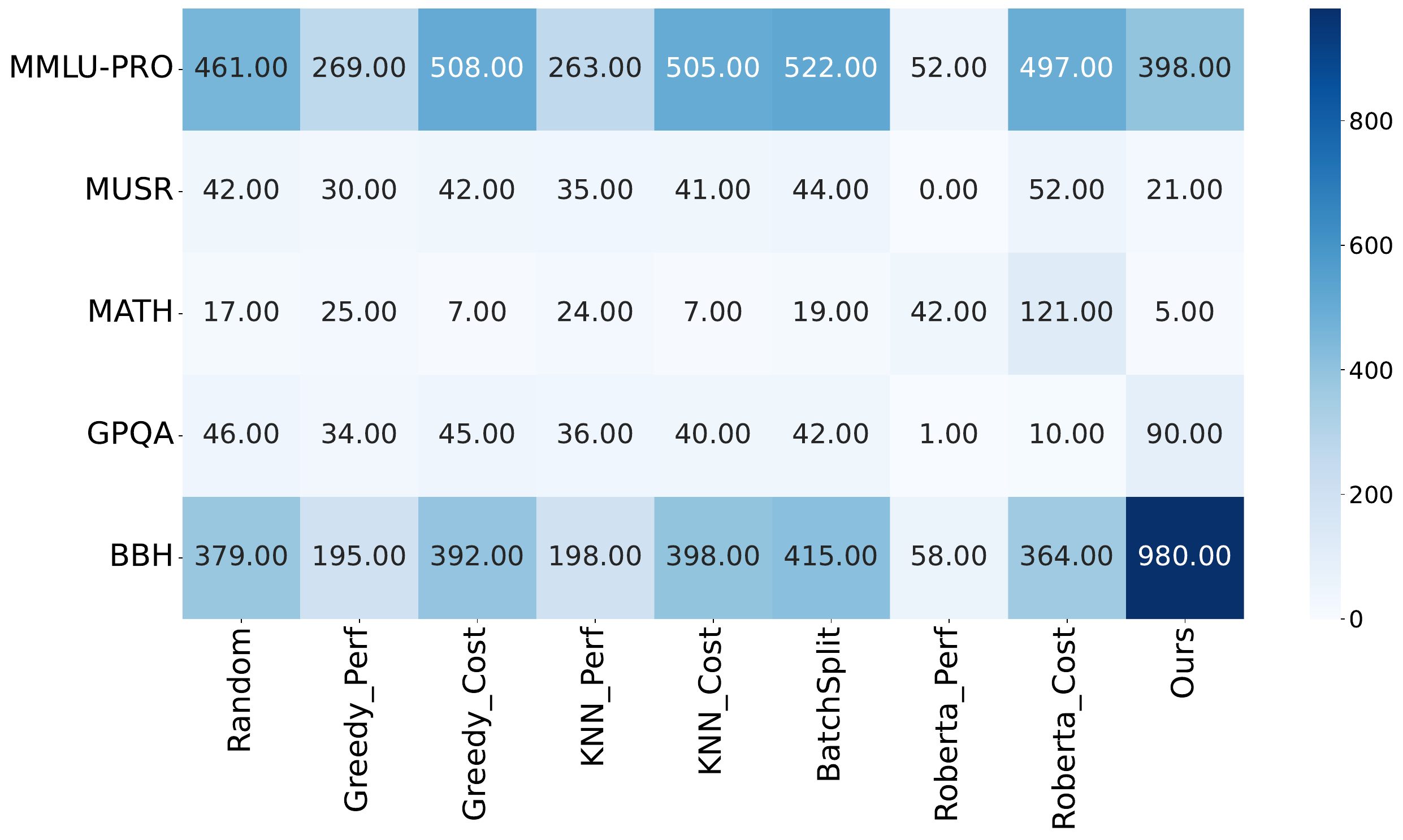}
     \caption{ Results across different query types in Open LLM Leaderboard v2.
    }
    \label{fig:lheat}
\end{figure}
\subsection{Routing Analysis by Query Types}\label{app:query_types}
To gain deeper insights into how our algorithm routes different types of queries, we analyze its routing behavior across the three benchmarks.
We categorize query types based on their data sources, as each source is designed to evaluate different LLM capabilities.
As shown in Figure~\ref{fig:rheat}, \ref{fig:sheat}, and~\ref{fig:lheat}, our algorithm consistently achieves superior or competitive routing performance in most query types across benchmarks.
These results demonstrate the generalizability and effectiveness of our algorithm in handling diverse query types, which indicates that its performance is not dependent on a narrow subset of tasks.

\section{Query-Traffic Considerations in LLM Routing}\label{app:dis}
In the practical LLM-serving systems, an additional important routing factor is query traffic patterns. 
However, the primary goal of this work is to develop a cost-efficient online routing algorithm, a problem that has gained increasing attention in recent literature~\cite{ong2025routellmlearningroutellms, hu2024routerbench, somerstep2025carrot}, especially given the high computational cost of LLMs. 
To isolate this core algorithmic problem, we adopt a standard assumption widely used in the online algorithms literature~\cite{devanur2009adwords, mehta2007adwords, goel2008online}: incoming queries arrive in a randomly permuted order. 
This assumption, discussed in detail in Appendix~\ref{app:assump}, allows us to abstract away complex traffic dynamics and focus on the fundamental cost-efficiency routing problem.

Under this assumption, our algorithm is designed to operate on \textbf{query-intrinsic features} that are independent of time and query order, such as response performance score and cost per query across the models. 
In contrast, \textbf{operational-state features}, such as current queue lengths or system load, are beyond the scope of our current formulation and are more aligned with a related but orthogonal problem: load balancing under traffic constraints.

Still, we believe our algorithm offers a natural foundation for incorporating such traffic-aware considerations. 
One potential approach is to discretize time into short windows and track the per-model backlog of unserved workload. 
Within each time window, a corresponding "time-bucket" MILP can be solved with additional traffic constraints, and any unserved workload can be carried over to the next window. 
For instance, let ${\ell}_{ij}$ denote the service time of query $j$ on LLM $i$, $b_{it}$ be the backlog of unserved workload for LLM $i$ at the start of time window $t$, and $A_{it}$ denote the available service capacity for LLM $i$ during window $t$. 
Then, for each window $t$ and LLM $i$, we can add the following constraint to the MILP: 
\begin{equation}
    \sum_j {\ell}_{ij} x_{ij} \leq \max\{0, A_{it} - b_{it}\}, \quad \forall i,t
\end{equation}
This ``time-bucket'' formulation helps smooth bursts and reduce queuing delays.
We leave the integration of such traffic-aware routing mechanisms to future work.

\section{Extended Related Work}\label{app:related}
\noindent\textbf{LLM Routing.}
Existing LLM routing research primarily falls into two paradigms.
The first focuses on improving response quality while managing cost, typically through ensembling outputs from different LLMs~\cite{jiang2023llm, wang2023fusing} or using cascading strategies that query LLMs sequentially based on their capabilities~\cite{chen2023frugalgpt, aggarwal2024automix, yue2023large, lee2023orchestrallm}. 
For instance, LLM-Blender~\cite{jiang2023llm} proposes an ensembling framework that combines outputs from multiple open-source LLMs and selects the optimal response.
~\cite{wang2023fusing}using outputs
Wang et al.~\cite{wang2023fusing} propose to fuse outputs of expert models that capture complementary aspects of the data distribution, to generate the final answer.
Frugal-GPT~\cite{chen2023frugalgpt} adopts a sequential querying strategy, invoking LLMs in order of increasing capability until a satisfactory response is obtained.
AutoMix~\cite{aggarwal2024automix} first uses a smaller model to self-verify the quality of the response, based on which it dynamically selects a suitably sized model for handling the query.
Although these approaches can improve performance, they incur high latency and computational cost due to multiple model invocations per query.

The second paradigm leverages historical or auxiliary datasets to estimate the performance and cost of each query and select the optimal LLM accordingly~\cite{ong2025routellmlearningroutellms, somerstep2025carrot, hu2024routerbench, ding2024hybrid, lu2023routing, jitkrittum2025universal, shnitzer2023large, stripelis2024tensoropera, feng2024graphrouter, hari2023tryage, wang2025mixllm}.
In this line of work, several approaches train model-based predictors using these labeled historical datasets.
For example, RouterLLM~\cite{ong2025routellmlearningroutellms} trains a BERT-base or causal LLM on the historical data annotated with human preferences to enhance routing accuracy. 
CARROT~\cite{somerstep2025carrot} trains Roberta-base models to estimate both performance and cost for each query, enabling selection of the optimal LLM under any user-defined trade-off between quality and cost.
HybridLLM~\cite{ding2024hybrid} adopts the synthetic preference labels from the MixInstruct dataset~\cite{jiang2023llm} based on BARTScore~\cite{yuan2021bartscore}, and trains a single BERT-based router for query routing.
Zooter~\cite{lu2023routing} distills reward signals from training queries and applies tag-based label enhancement to train a routing function that facilitates expert selection.
TensorOpera~\cite{stripelis2024tensoropera} introduces a soft label-based strategy based on the BERT similarity scores to train the BERT-based predictor for query routing.
GraphRouter~\cite{feng2024graphrouter} constructs a heterogeneous graph to capture the contextual relationship between the query requirements and the LLM capabilities for informed routing decisions.
While these training-based approaches are effective for certain LLM deployment settings, they introduce nontrivial training overhead. 
Furthermore, adapting them to varying LLM deployment configurations typically requires retraining, which makes them unsuitable for resource-constrained online routing with diverse LLM deployment configurations.

Some methods avoid training by using approaches such as KNN~\cite{hu2024routerbench} or similarity-weighted (SW) ranking~\cite{ong2025routellmlearningroutellms} to estimate the performance and cost based on the historical data.
While these approaches avoid model training overhead, they still incur substantial computational overhead and latency, as they rely on brute-force comparisons against the entire historical dataset to retrieve similar examples or compute similarity scores without any optimization.
Specifically, they operate with a high search complexity of $O(N)$, where $N$ is the size of the dataset.
As a result, these methods are difficult to scale and impractical for high-volume, budget-constrained online routing scenarios.

Recent efforts~\cite{vsakota2024fly, mohammadshahi2024routoo, rama2025cerebrum} have explored formulating LLM routing as MILP.
For instance, Sakota et al.~\cite{vsakota2024fly} introduce two MILP-based strategies, targeting performance-oriented and cost-oriented, respectively. 
However, these methods face significant challenges in high-volume online routing settings, where queries arrive sequentially, rather than simultaneously, making the offline optimal infeasible to compute.

We complement these works by introducing the first training-free and efficient online routing method with provable performance guarantees, tailored for high-volume, budget-constrained, and dynamic LLM-serving environments.

\noindent\textbf{Online Matching.}
This problem has been extensively studied for decades, with much of the literature focusing on important special cases of the general online submodular welfare maximization problem~\cite{kapralov2013online, korula2015online}. 
These include the classical unweighted online bipartite matching problem~\cite{kalyanasundaram2000optimal, karp1990optimal, kesselheim2013optimal}, its extension to vertex-weighted matching~\cite{buchbinder2007online, aggarwal2011online}, the display ads allocation problem~\cite{feldman2009online, chen2011real}, and the AdWords problem~\cite{mehta2007adwords, devanur2009adwords, buchbinder2007online}.

\section{Broader Impacts}\label{app:imp}
Our algorithm proposes an efficient, training-free solution for online routing in high-volume, budget-constrained LLM serving. 
It significantly reduces serving costs and improves performance, which enables more sustainable, scalable, and cost-effective multi-LLM deployment.
As a technical contribution, it does not pose direct societal risks. 
We believe it promotes broader accessibility to LLM services, and future work can further address potential concerns by incorporating responsible use guidelines.

%% file: tables/benchmark.tex
\begin{table*}[ht]
\small
\caption{Benchmark details for RouterBench, SPROUT, and Open LLM Leaderboard v2 in our experiments.}
\label{tab:benchmark}
\centering
\begin{minipage}[t]{0.33\textwidth}
\centering
\caption*{(a) RouterBench}
\begin{tabular}{lc}
\toprule
\textbf{Dataset} & \textbf{Size} \\
\midrule
MMLU~\cite{hendrycks2020measuring} & 14042 \\
Hellaswag~\cite{zellers2019hellaswag} & 10042 \\
GSM8K~\cite{cobbe2021training} & 7450 \\
ARC Challenge~\cite{clark2018think} & 1470 \\
Winogrande~\cite{sakaguchi2021winogrande} & 1267 \\
MBPP~\cite{austin2021program} & 427 \\
MT-Bench~\cite{NEURIPS2023_91f18a12} & 80 \\
Chinese & 785 \\
Consensus Summary & 362 \\
Bias Detection & 285 \\
Test Match & 3 \\
Accounting Audit & 30 \\
Abstract2title & 254 \\
\textbf{Total} & 36497 \\
\bottomrule
\end{tabular}
\end{minipage}%
\hfill
\begin{minipage}[t]{0.33\textwidth}
\centering
\caption*{(b) SPROUT}
\begin{tabular}{lc}
\toprule
\textbf{Dataset} & \textbf{Size} \\
\midrule
MATH Lvl 1-5~\cite{hendrycks2021measuring} & 9884  \\
MMLU-PRO~\cite{wang2024mmlu} & 11786  \\
GPQA~\cite{rein2024gpqa} & 541 \\
MUSR~\cite{sprague2023musr} & 748 \\
RAGBench~\cite{friel2024ragbench} & 1827  \\
Openhermes 2.5~\cite{openhermes} & 19455  \\
\textbf{Total} & 44241 \\
-- & --\\
-- & --\\
-- & --\\
-- & --\\
-- & --\\
-- & --\\
-- & --\\
\bottomrule
\end{tabular}
\end{minipage}
\hfill
\begin{minipage}[t]{0.33\textwidth}
\centering
\caption*{(c) Open LLM Leaderboard v2}
\begin{tabular}{lc}
\toprule
\textbf{Dataset} & \textbf{Size} \\
\midrule
MMLU-PRO~\cite{wang2024mmlu} & 12032 \\
MUSR~\cite{sprague2023musr} &  756\\
MATH Lvl 1-5~\cite{hendrycks2021measuring} & 1324 \\
GPQA~\cite{rein2024gpqa} & 1192 \\
BBH~\cite{suzgun2022challengingbigbenchtaskschainofthought} & 5761 \\
\textbf{Total} &  21065\\
-- & --\\
-- & --\\
-- & --\\
-- & --\\
-- & --\\
-- & --\\
-- & --\\
-- & --\\
\bottomrule
\end{tabular}
\end{minipage}%
\end{table*}

%% file: tables/cost.tex
\begin{table*}[t]
\scriptsize
\caption{Model lists and token costs (\$/1M tokens) for RouterBench(left), SPROUT (middle), and Open LLM Leaderboard v2 (right).}
\label{tab:combined_model_costs}
\centering
\begin{minipage}[t]{0.2\textwidth}
\centering
\caption*{(a) RouterBench}
\begin{tabular}{l}
\toprule
\textbf{Models}  \\
\midrule
WizardLM-13B-V1.2 \\
claude-instant-v1 \\
claude-v1 \\
claude-v2 \\
gpt-3.5-turbo-1106 \\
gpt-4-1106-preview \\
code-llama-instruct-34b \\
llama-2-70b-chat \\
mistral-7b-chat \\
mixtral-8x7b-chat \\
Yi-34B-Chat \\
-- \\
-- \\
-- \\
-- \\
-- \\
-- \\
-- \\
\bottomrule
\end{tabular}
\end{minipage}%
\hfill
\begin{minipage}[t]{0.45\textwidth}
\centering
\caption*{(b) SPROUT}
\begin{tabular}{lcc}
\toprule
\textbf{Models} & \textbf{Input Cost} & \textbf{Output Cost} \\
\midrule
claude-3-5-sonnet-v1 & 3.00 & 15.00 \\
titan-text-premier-v1 & 0.50 & 1.50 \\
openai-gpt-4o & 2.50 & 10.00 \\
openai-gpt-4o-mini & 0.15 & 0.60 \\
granite-3-2b-instruct & 0.10 & 0.10 \\
granite-3-8b-instruct & 0.20 & 0.20 \\
llama-3-1-70b-instruct & 0.90 & 0.90 \\
llama-3-1-8b-instruct & 0.20 & 0.20 \\
llama-3-2-1b-instruct & 0.06 & 0.06 \\
llama-3-2-3b-instruct & 0.06 & 0.06 \\
llama-3-3-70b-instruct & 0.90 & 0.90 \\
mixtral-8x7b-instruct & 0.60 & 0.60 \\
llama-3-405b-instruct & 3.50 & 3.50 \\
-- & -- & --\\
-- & -- & --\\
-- & -- & --\\
-- & -- & --\\
-- & -- & --\\
\bottomrule
\end{tabular}
\end{minipage}
\hfill
\begin{minipage}[t]{0.33\textwidth}
\centering
\caption*{(c) Open LLM Leaderboard v2}
\begin{tabular}{lc}
\toprule
\textbf{Models} & \textbf{Cost} \\
\midrule
Mixtral-8x7B-DPO & 0.6 \\
Yi-34B-Chat & 0.8 \\
QwQ-32B-Preview & 1.2 \\
Qwen2-72B-Instruct & 0.9 \\
Qwen2.5-7B-Instruct & 0.3 \\
Qwen2.5-72B-Instruct & 1.2 \\
WizardLM-2-8x22B & 1.2 \\
deepseek-llm-67b-chat & 0.9 \\
gemma-2-27b-it & 0.8 \\
gemma-2-9b-it & 0.3 \\
gemma-2b-it & 0.1 \\
Llama-2-13b & 0.3 \\
Meta-Llama-3.1-70B & 0.9 \\
Mistral-7B-Instruct-v0.1 & 0.2 \\
Mistral-7B-Instruct-v0.2 & 0.2 \\
Mistral-7B-Instruct-v0.3 & 0.2 \\
Mixtral-8x7B-Instruct-v0.1 & 0.6 \\
Llama-3.1-Nemotron-70B & 0.9 \\
\bottomrule
\end{tabular}
\end{minipage}
\end{table*}

%% file: tables/perf-cost.tex
\begin{table}[t]
\small
\centering
\caption{Average cost and performance of models in RouterBench on historical data.}
\label{tab:routerbench_perf_cost}
\begin{tabular}{lcccc}
\toprule
\textbf{Model} & \textbf{Cost} & \textbf{Perf} & \textbf{Perf/Cost} & \textbf{(Perf/Cost)$^{0.5}$}\\
\midrule
WizardLM-13B-V1.2                & 7.27e\textminus05 & 0.432 & 5944 & 77.10 \\
claude-instant-v1                         & 2.32e\textminus04 & 0.598 & 2581 & 50.80 \\
claude-v1                                 & 2.14e\textminus03 & 0.631 &   295 & 17.18 \\
claude-v2                                 & 2.41e\textminus03 & 0.636 &   264 & 16.24 \\
gpt-3.5-turbo-1106                        & 2.42e\textminus04 & 0.617 & 2546 & 50.45 \\
gpt-4-1106-preview                        & 3.28e\textminus03 & 0.781 &   238 & 15.43 \\
code-llama-instruct-34b         & 1.71e\textminus04 & 0.203 & 1182 & 34.38 \\
llama-2-70b-chat                     & 2.02e\textminus04 & 0.328 & 1627 & 40.34 \\
mistral-7b-chat                 & 4.56e\textminus05 & 0.308 & 6768 & 82.27 \\
mixtral-8x7b-chat               & 1.34e\textminus04 & 0.550 & 4098 & 64.01 \\
Yi-34B-Chat                   & 1.85e\textminus04 & 0.648 & 3503 & 59.19 \\
\bottomrule
\end{tabular}
\end{table}

\begin{table}[t]
\small
\centering
\caption{Average cost and performance of models in SPROUT on historical data.}
\label{tab:sprout_perf_cost}
\begin{tabular}{lcccc}
\toprule
\textbf{Model} & \textbf{Cost} & \textbf{Perf} & \textbf{Perf/Cost} & \textbf{(Perf/Cost)$^{0.5}$}\\
\midrule
claude-3-5-sonnet-v1                 & 7.65e\textminus03 & 0.827 &     108 & 10.39 \\
titan-text-premier-v1                & 5.64e\textminus04 & 0.579 &   1027 & 32.04 \\
openai-gpt-4o                            & 4.92e\textminus03 & 0.846 &     172 & 13.11 \\
openai-gpt-4o-mini                       & 3.40e\textminus04 & 0.808 &   2378 & 48.76 \\
granite-3-2b-instruct            & 8.54e\textminus05 & 0.553 &   6473 & 80.46 \\
granite-3-8b-instruct        & 1.50e\textminus04 & 0.659 &   4403 & 66.36 \\
llama-3-1-70b-instruct              & 7.17e\textminus04 & 0.810 &   1130 & 33.62 \\
llama-3-1-8b-instruct               & 2.43e\textminus04 & 0.690 &   2838 & 53.27 \\
llama-3-2-1b-instruct               & 6.67e\textminus05 & 0.460 &   6904 & 83.09 \\
llama-3-2-3b-instruct               & 6.47e\textminus05 & 0.629 &   9722 & 98.60 \\
llama-3-3-70b-instruct              & 5.52e\textminus04 & 0.804 &   1457 & 38.17 \\
llama-3-405b-instruct               & 2.01e\textminus03 & 0.776 &     385 & 19.63 \\
mixtral-8x7b-instruct              & 3.74e\textminus04 & 0.616 &   1648 & 40.59 \\
\bottomrule
\end{tabular}
\end{table}

\begin{table}[t]
\small
\centering
\caption{Average cost and performance of models in Open LLM Leaderboard v2 on historical data.}
\label{tab:leaderboard_perf_cost}
\begin{tabular}{lcccc}
\toprule
\textbf{Model} & \textbf{Cost} & \textbf{Perf} & \textbf{Perf/Cost} & \textbf{(Perf/Cost)$^{0.5}$}\\
\midrule
Yi-34B-Chat                                & 6.57e\textminus04 & 0.428 &   652 & 25.53 \\
Mixtral-8x7B-DPO      & 4.78e\textminus04 & 0.401 &   839 & 28.97 \\
QwQ-32B-Preview                              & 8.90e\textminus04 & 0.552 &   621 & 24.91 \\
Qwen2-72B-Instruct                           & 6.67e\textminus04 & 0.562 &   842 & 29.02 \\
Qwen2.5-72B-Instruct                         & 8.90e\textminus04 & 0.561 &   630 & 25.10 \\
Qwen2.5-7B-Instruct                          & 2.22e\textminus04 & 0.420 & 1887 & 43.44 \\
WizardLM-2-8x22B                        & 9.85e\textminus04 & 0.491 &   499 & 22.34 \\
deepseek-llm-67b-chat                 & 7.05e\textminus04 & 0.413 &   586 & 24.21 \\
gemma-2-27b-it                             & 6.13e\textminus04 & 0.462 &   753 & 27.43 \\
gemma-2-9b-it                              & 2.30e\textminus04 & 0.419 & 1826 & 42.72 \\
gemma-2b-it                                & 7.66e\textminus05 & 0.191 & 2489 & 49.89 \\
Llama-2-13b                           & 2.47e\textminus04 & 0.227 &   919 & 30.31 \\
Meta-Llama-3.1-70B            & 6.44e\textminus04 & 0.548 &   852 & 29.18 \\
Mistral-7B-Instruct-v0.1                & 1.43e\textminus04 & 0.258 & 1806 & 42.50 \\
Mistral-7B-Instruct-v0.2                & 1.64e\textminus04 & 0.311 & 1894 & 43.52 \\
Mistral-7B-Instruct-v0.3                & 1.64e\textminus04 & 0.336 & 2044 & 45.21 \\
Mixtral-8x7B-Instruct-v0.1              & 4.92e\textminus04 & 0.379 &   770 & 27.74 \\
nvidia/Llama-3.1-Nemotron-70B       & 7.39e\textminus04 & 0.506 &   686 & 26.19 \\
\bottomrule
\end{tabular}
\end{table}

%% file: figures_tex/basedata.tex
\begin{figure}[t]
    \centering
    \includegraphics[width=\linewidth]{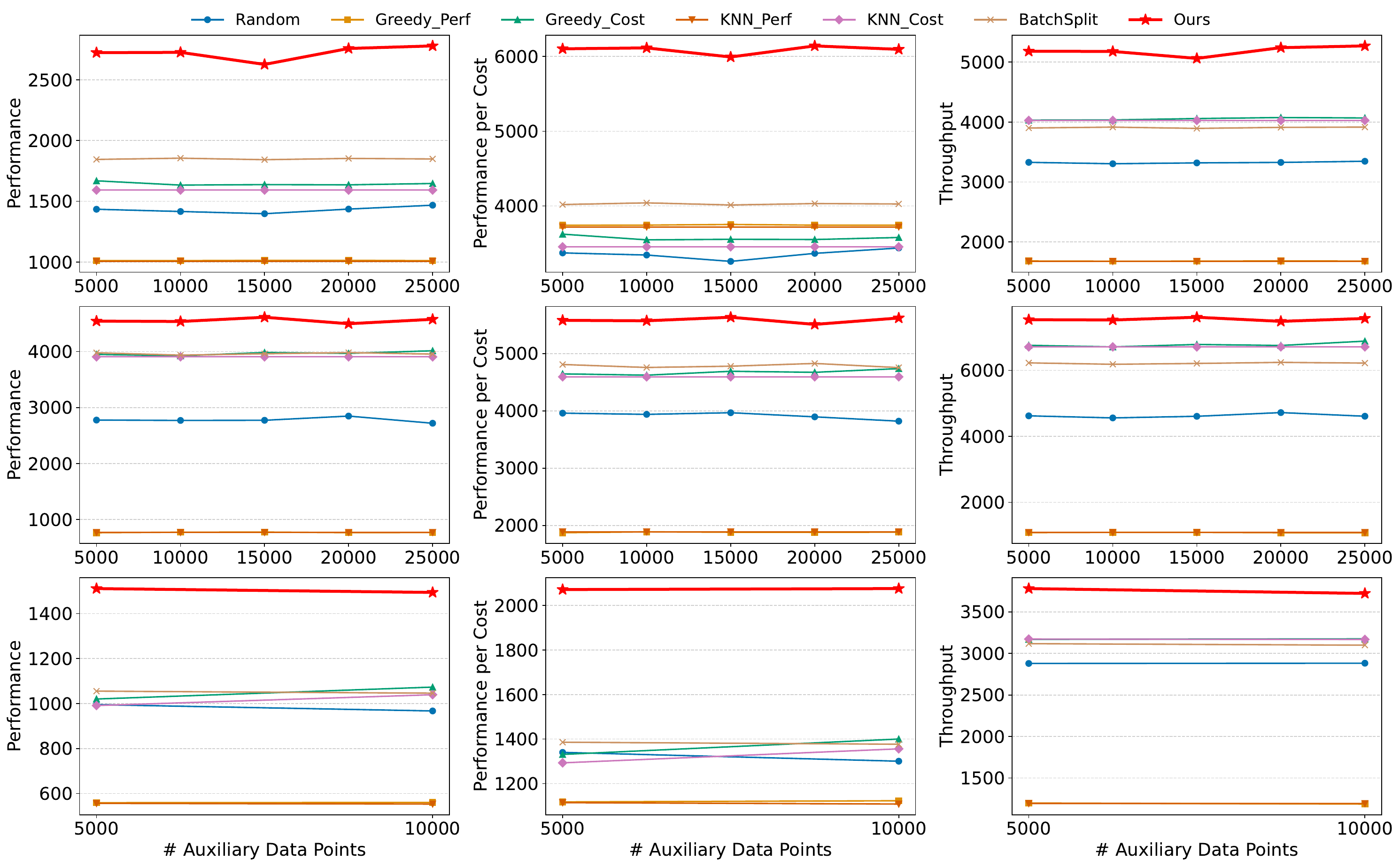}
    \caption{Results when varying the number of historical data points. 
    Rows correspond to different datasets: RouterBench (top), SPROUT (middle), and
    Open LLM Leaderboard v2 (bottom).
    }
    \label{fig:basesize}
\end{figure}

%% file: figures_tex/topk.tex
\begin{figure}[ht]
    \centering
    \includegraphics[width=\linewidth]{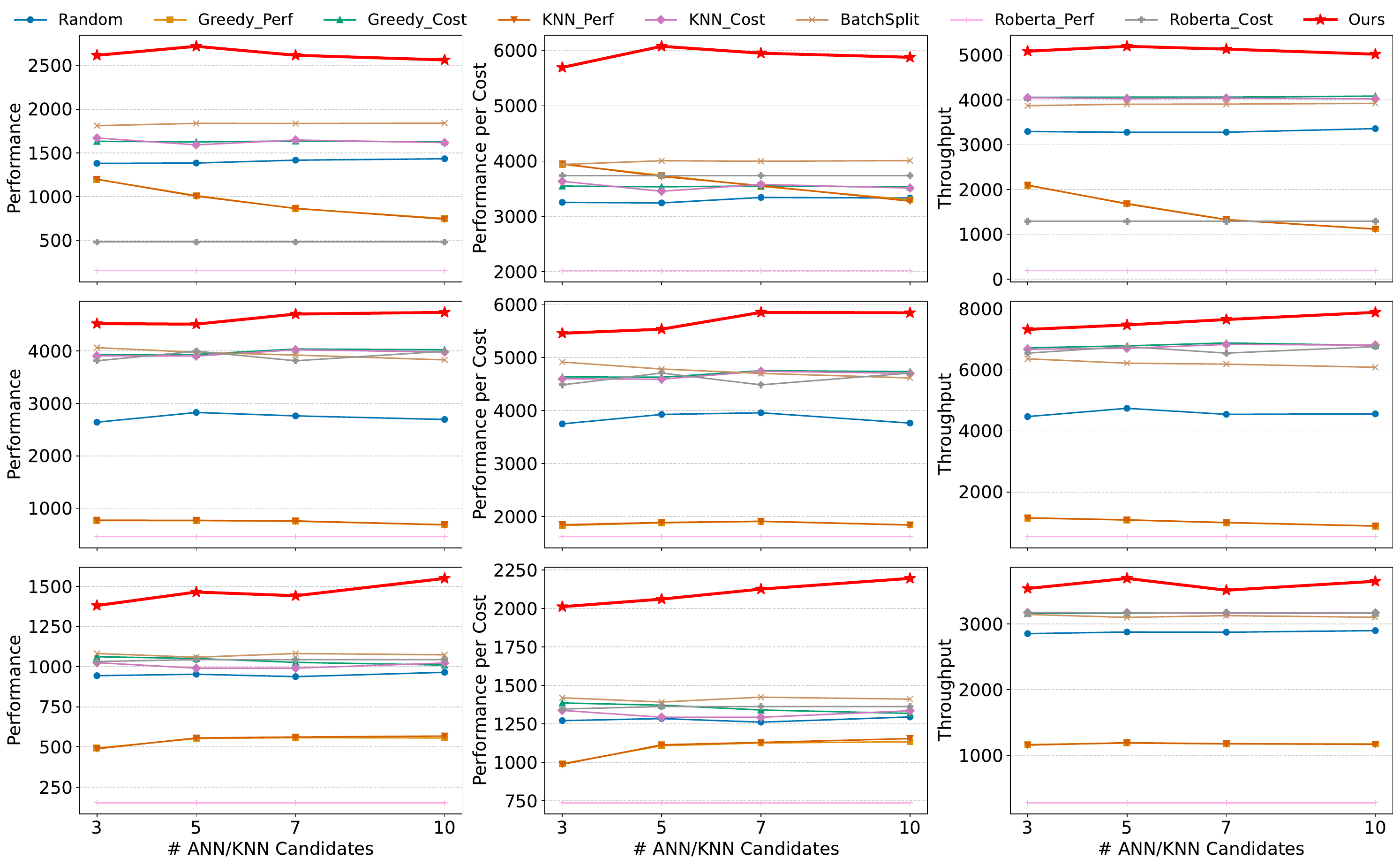}
    \caption{Results when varying the number of search candidates. 
    Rows correspond to different datasets: RouterBench (top), SPROUT (middle), and Open LLM Leaderboard v2
    (bottom).
    }
    \label{fig:topk}
\end{figure}

%% file: figures_tex/embed.tex
\begin{figure}[t]
    \centering
    \includegraphics[width=\linewidth]{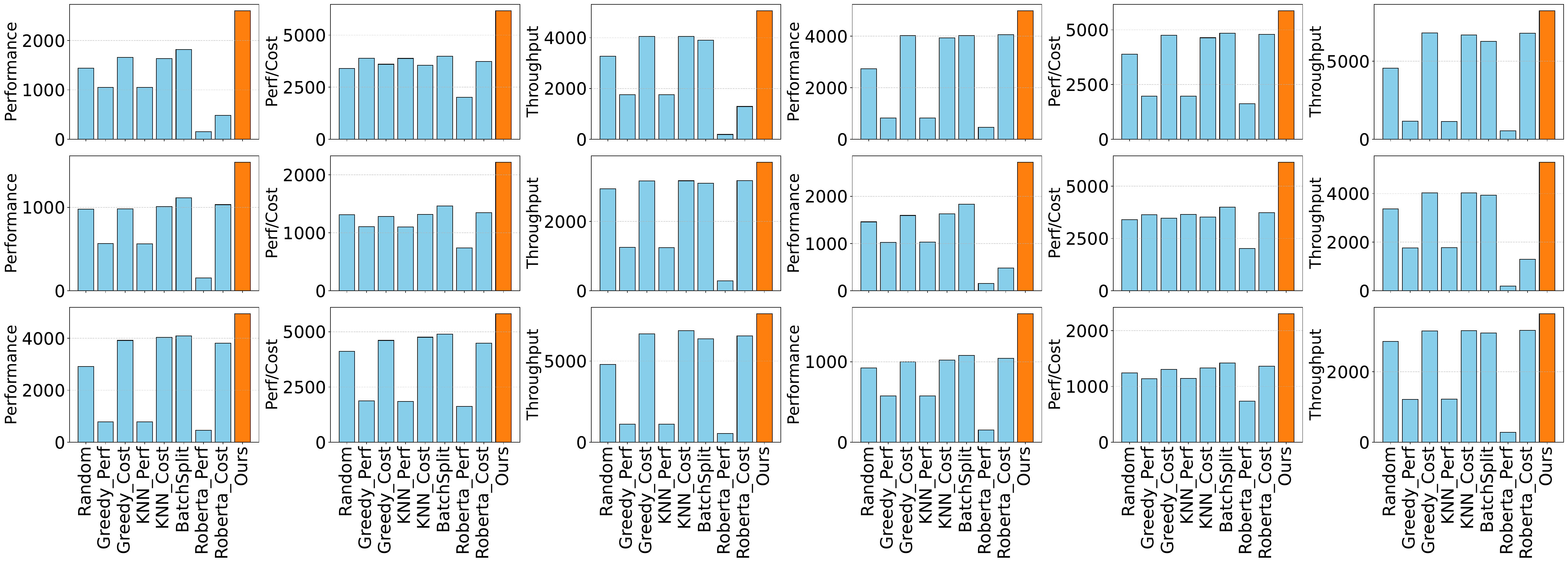}
    \caption{Results using different embedding models. Each group of 9 subfigures corresponds
    to one embedding: (a) gte-Qwen2-1.5B-instruct (subfigures 1–9), (b) SFR-Embedding-2\_R (subfigures 10–18). Within each group, the first three subfigures correspond to RouterBench, the next three to SPROUT, and the last three to Open LLM Leaderboard v2.
    }
    \label{fig:embed}
\end{figure}

%% file: figures_tex/alpha.tex
\begin{figure}[t]
    \centering
    \includegraphics[width=\linewidth]{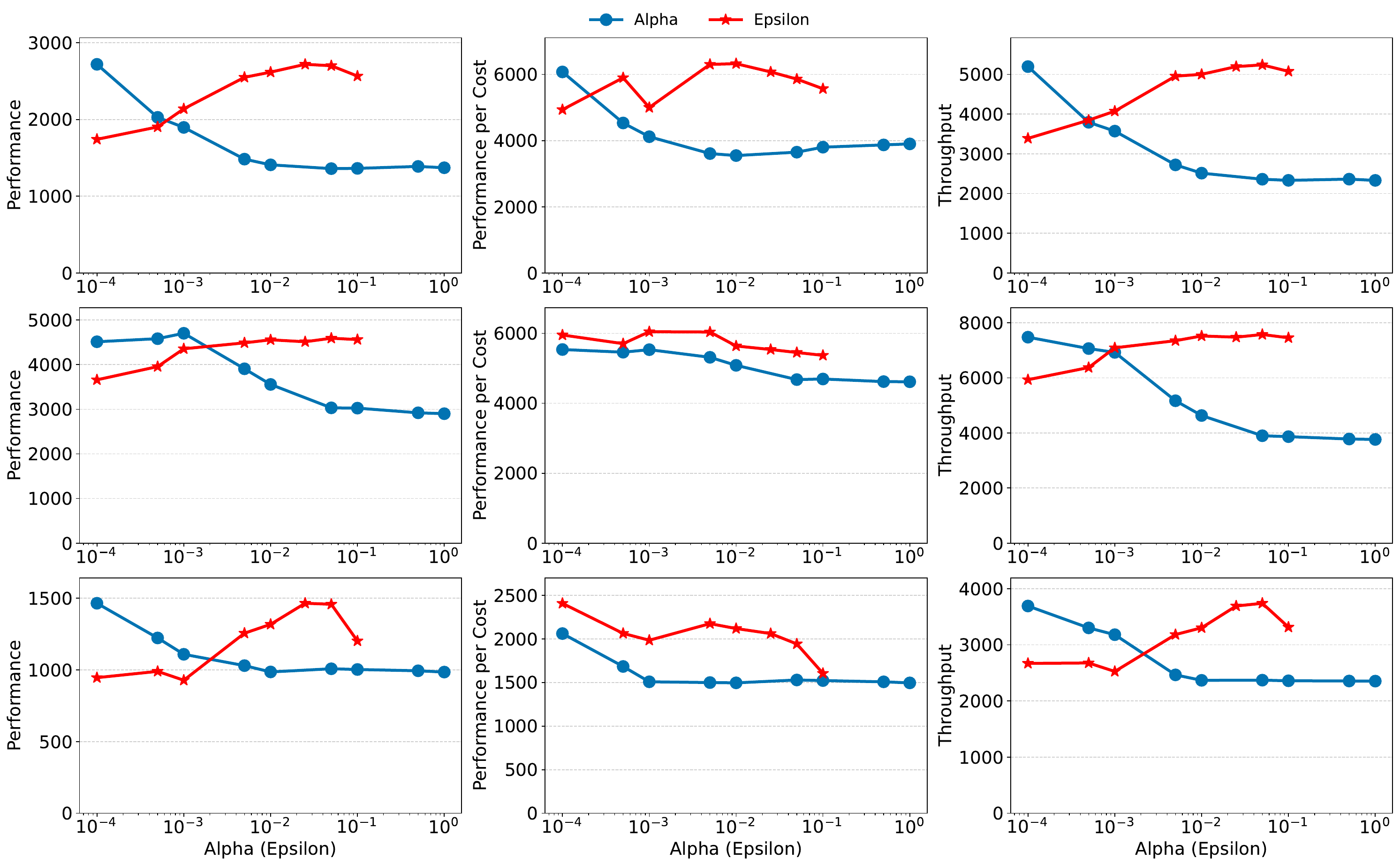}
     \caption{ Results when varying $\alpha$ and $\epsilon$. Rows
    correspond to different datasets: RouterBench (top), SPROUT (middle), and Open LLM Leaderboard v2 (bottom).
}
    \label{fig:alpha_eps}
\end{figure}